\pdfoutput=1

\documentclass[prl,twocolumn,floatfix,showpacs,superscriptaddress]{revtex4}

\usepackage{graphicx}
\usepackage{natbib}
\usepackage{epstopdf}
\usepackage{amsmath}
\usepackage{amssymb}
\usepackage{mathbbol}
\usepackage{xcolor}
\usepackage{lipsum}
\usepackage{amssymb}

\newcommand{\beq}{\begin{equation}}
\newcommand{\eeq}{\end{equation} \smallskip}
\newcommand{\beqy}{\begin{eqnarray}}
\newcommand{\eeqy}{\end{eqnarray} \smallskip}

\newcommand{\bit}{\begin{itemize}}
\newcommand{\eit}{\end{itemize}}
\newcommand{\bmat}{\begin{pmatrix}}
\newcommand{\emat}{\end{pmatrix}}
\newcommand{\red}[1]{\textcolor{red}{#1}}

\newcommand{\vect}[1]{{\mathbf #1}}
\newcommand{\Frac}[2]{\displaystyle\frac{#1}{#2}}

\begin{document}

\title{Dynamical critical exponents in driven-dissipative quantum systems}

\author{P. Comaron}
\address{Joint Quantum Centre (JQC) Durham-Newcastle, School of Mathematics, Statistics and Physics,
Newcastle University, Newcastle upon Tyne, NE1 7RU, United Kingdom}

\author{G. Dagvadorj}
\address{Department of Physics and Astronomy, University College London,
Gower Street, London, WC1E 6BT, United Kingdom}
\address{Department of Physics, University of Warwick, Coventry, CV4 7AL, United Kingdom}

\author{A. Zamora}
\address{Department of Physics and Astronomy, University College London,
Gower Street, London, WC1E 6BT, United Kingdom} 

\author{I. Carusotto}
\address{INO-CNR BEC Center and Universit\`a di Trento, via Sommarive 14, I-38123 Povo, Italy
} 

\author{N. P. Proukakis}
\address{Joint Quantum Centre (JQC) Durham-Newcastle, School of Mathematics, Statistics and Physics,
Newcastle University, Newcastle upon Tyne, NE1 7RU, United Kingdom}

\author{M. H. Szyma\'nska}
\address{Department of Physics and Astronomy, University College London,
Gower Street, London, WC1E 6BT, United Kingdom}
\email{m.szymanska@ucl.ac.uk}

\date{\today}

\begin{abstract}
We study the phase-ordering of parametrically and incoherently driven microcavity polaritons after an infinitely rapid quench across the critical region. We confirm that the system, despite its driven-dissipative nature, fulfils dynamical scaling hypothesis for both driving schemes by exhibiting self-similar patterns for the two-point correlator at late times of the phase ordering. We show that polaritons are characterised by the dynamical critical exponent $z \approx 2$ with topological defects playing a fundamental role in the dynamics, giving logarithmic corrections both to the power-law decay of the number of vortices and to the associated growth of the characteristic length-scale.

\end{abstract}


\maketitle


Two-dimensional (2d) quantum driven-dissipative systems, such as
exciton-polaritons in microcavities \red{\cite{carusotto2013quantum}},
display rich universal critical phenomena due to the interplay between
drive, dissipation and potential anisotropy and their collective
dynamics induced by the Bose-degeneracy
\red{\cite{sieberer2016keldysh}}.  The picture is already quite
complex even at the level of the steady-state, where the system has
been predicted to adopt Kardar-Parisi-Zhang (KPZ)
\red{\cite{kardar1986dynamic}} or Berezinskii-Kosterlitz-Thouless (BKT)
\red{\cite{kosterlitz1973ordering}} type scaling, depending on the
subtle interplay between drive and dissipation, and finite size and
effective anisotropy
\red{\cite{altman2015twodimensional,zamora2016driving}}. Moreover, the even more
advanced question of how continuous drive and dissipation affect
dynamical critical behaviour remains largely open.

In this vein, one of the most useful concepts is the scaling
hypothesis. Complex systems at criticality display self-similar
patterns, since the system is statistically equivalent after a global
and arbitrary change of scale \red{\cite{bray2002theory}}. The scaling
phenomenon also reveals the existence of universal critical exponents,
characterising macroscopic properties of the system at critical
points.  Such critical exponents for driven-dissipative microcavity
polaritons have not been measured to date while theoretical
predictions are subject to debate.  Specifically, 2d
driven-dissipative systems described by KPZ-like phase dynamics are
expected to show the dynamical critical exponent $z\approx1.61$ while
in the equilibrium limit, where the KPZ nonlinearity ceases to be
important, we expect $z=2$.  Moreover, it has been suggested that the
dynamical critical exponent $z$ for microcavity polaritons coupled to a reservoir
takes values of either $1$ or $2$, depending on system parameters
\red{\cite{kulczykowski2017phase}}, putting in question the
universality of the phase ordering in this system.

In this letter, we explore the critical properties of 2d-microcavity polaritons, and in particular the characteristic length and
density of topological defects (vortices), by
studying the phase ordering (scaling) dynamics after an infinitely
rapid quench from the disordered to deep in the (quasi)ordered
\red{\cite{hohenberg1977theory,bray2002theory}} phase. 
In order to capture the universal properties of the phase ordering
process we study the polariton system under three different pumping
configurations, focusing on experimentally realistic conditions. Specifically we study the coherent pumping in the optical parametric oscillator (OPO)
regime, and the incoherent pumping (IP) with and without a
frequency-selective pumping mechanism.  In particular, we report
strong numerical evidence that the three different polariton systems
all show the same critical behaviour, characterised by the dynamical
critical exponent $z\approx 2$ with logarithmic correction to the
diffusive dynamics.
Thus, we find that the universal properties of the system, also in the
dynamical case, are dominated by BKT-type of physics in analogy to
the static case considered recently theoretically
\red{\cite{dagvadorj2015nonequilibrium}} and experimentally
\red{\cite{caputo2016topological}}. It seems that the KPZ nonlinear
terms do not play a role during dynamical crossing of
the critical point, at least for the realistic system sizes considered in this work.

\paragraph*{System and method. }
\label{sec:coherent_pump}

Our system consists of an ensemble of bosonic particles (excitons (X) and
photons (C) for {parametrically}-pumped, and lower polaritons (LP) for
incoherently-driven case) with finite lifetime interacting via contact
interactions in two dimensions, and driven in two distinct ways:
{parametrically} and incoherently.  The dynamical equations can be derived
using Keldysh field theory by including the classical fluctuations to
all orders, but quantum fluctuations to the second order, which is
appropriate in the long-wavelength limit, and employing the Martin-Siggia-Rose (MSR)
formalism (for review see \red{\cite{sieberer2016keldysh}}).  An alternative
derivation can be performed using Fokker-Planck equations for the
Wigner function truncated to the third-order
\red{\cite{carusotto2005spontaneous,carusotto2013quantum}}. Both methods lead to the same stochastic
equation for the field $\psi(\mathbf{r},t)$ with the noise term
accounting for quantum and thermal fluctuations.

For the parametrically driven case, the finite grid version with 
$\hbar=1$ reads  \red{\cite{dagvadorj2015nonequilibrium}}:
\begin{equation}
i d \begin{pmatrix} \psi_X \\ \psi_C \end{pmatrix} =
dt\left[H_{\mathrm{MF}} \begin{pmatrix} \psi_X \\ \psi_C \end{pmatrix}
+ \begin{pmatrix} 0 \\ F_p \end{pmatrix}\right] +
\begin{pmatrix} \sqrt{\kappa_X} \: dW_X \\ \sqrt{\kappa_C} \:
dW_C \end{pmatrix}\;,
\label{eq:wigner}
\end{equation}
where $\psi_{X,C}=\psi_{X,C}(\vect{r},t)$ (with $\vect{r}=(x,y)$) are
the exciton, and cavity-photon fields respectively, $dW_X$ and $dW_C$ are
the complex-valued zero-mean white Wiener noise terms with $\langle
dW^{*}_l (\vect{r},t) dW_m (\vect{r}',t) \rangle
=\delta_{\vect{r},\vect{r}'} \delta_{l,m}dt$. The external
monochromatic {coherent} pump $F_p= f_p e^{i (\vect{k}_p \cdot \vect{r}
- \omega_p t)}$ injects photons with momentum $\vect{k}_p$ and
frequency $\omega_p$ while both fields decay with their corresponding
rates $\kappa_X$ and $\kappa_C$: here we use a shorthand notation
\begin{displaymath}
H_{\mathrm{MF}}=
\begin{pmatrix} \frac{-\nabla^2}{2m_X} + g_X
(|\psi_{X}|^2-\frac{1}{dV}) - i\kappa_X & \frac{\Omega_R}{2} \\
\frac{\Omega_R}{2} & \frac{-\nabla^2}{2m_C} -i \kappa_C\end{pmatrix}
\; , 
\end{displaymath} 
with $m_X$ and $m_C$ ($= 2.3 \ 10^{-5} m_e$) the exciton and cavity-photon masses
respectively. Since $m_X \gg m_C$, we consider the limit $m_X \to \infty$ and, consequently, the exciton field kinetic energy term disappears from Eq. \eqref{eq:wigner}. The exciton-photon Rabi-splitting is given by $\Omega_R$,
the exciton-exciton interaction strength by $g_X$ and $dV=a^2$ is the numerical grid unit cell
area (lattice spacing $a$). We solve Eq. \eqref{eq:wigner} for parameters
typical of current OPO experiments
\red{\cite{sanvitto2010persistent,dagvadorj2015nonequilibrium,carusotto2013quantum}},
namely:
$\Omega_R\approx4.4meV$, 
$g_X\approx 2 \times 10^{-3} \ meV\mu m^2$.
We consider $\kappa_X=\kappa_C$, with $\kappa_C = 1/\tau_C$ and photon lifetime $\tau_C=6.58 ps$.
We set $\textbf{k}_p=(1.6,0)\mu m^{-1}$,  $\omega_p=\omega_{LP}(\textbf{k}_p)$ 
where
$\omega_{LP}$ is the lower polariton
dispersion.  Here, we present results for a grid of $512^2$ lattice points with
lattice spacing $a=0.87\mu m$ (total unit cell of $L_x=L_y=444.42\mu
m$). 
%
\begin{figure}
\includegraphics[width=1\linewidth]{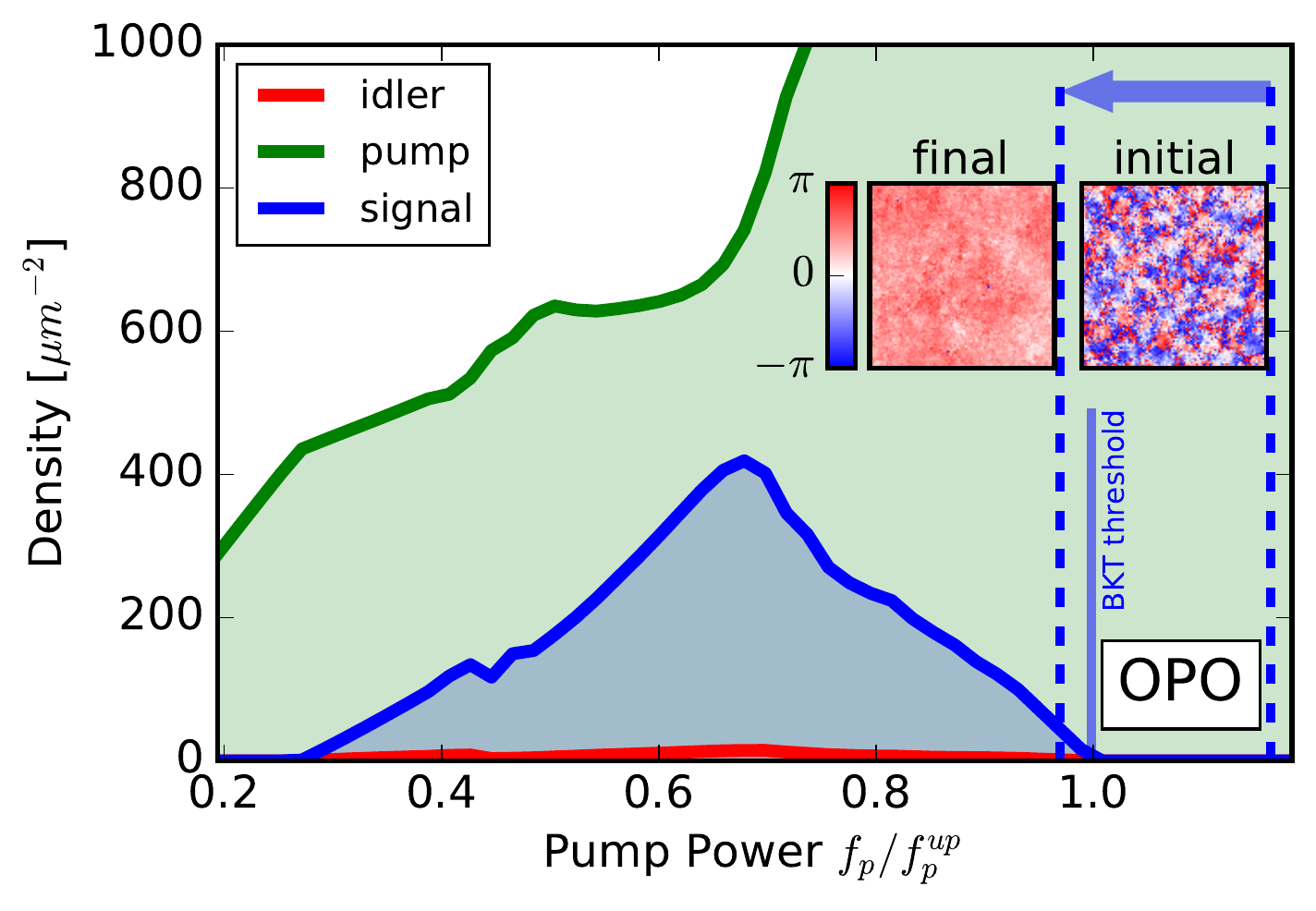}
\includegraphics[width=1\linewidth]{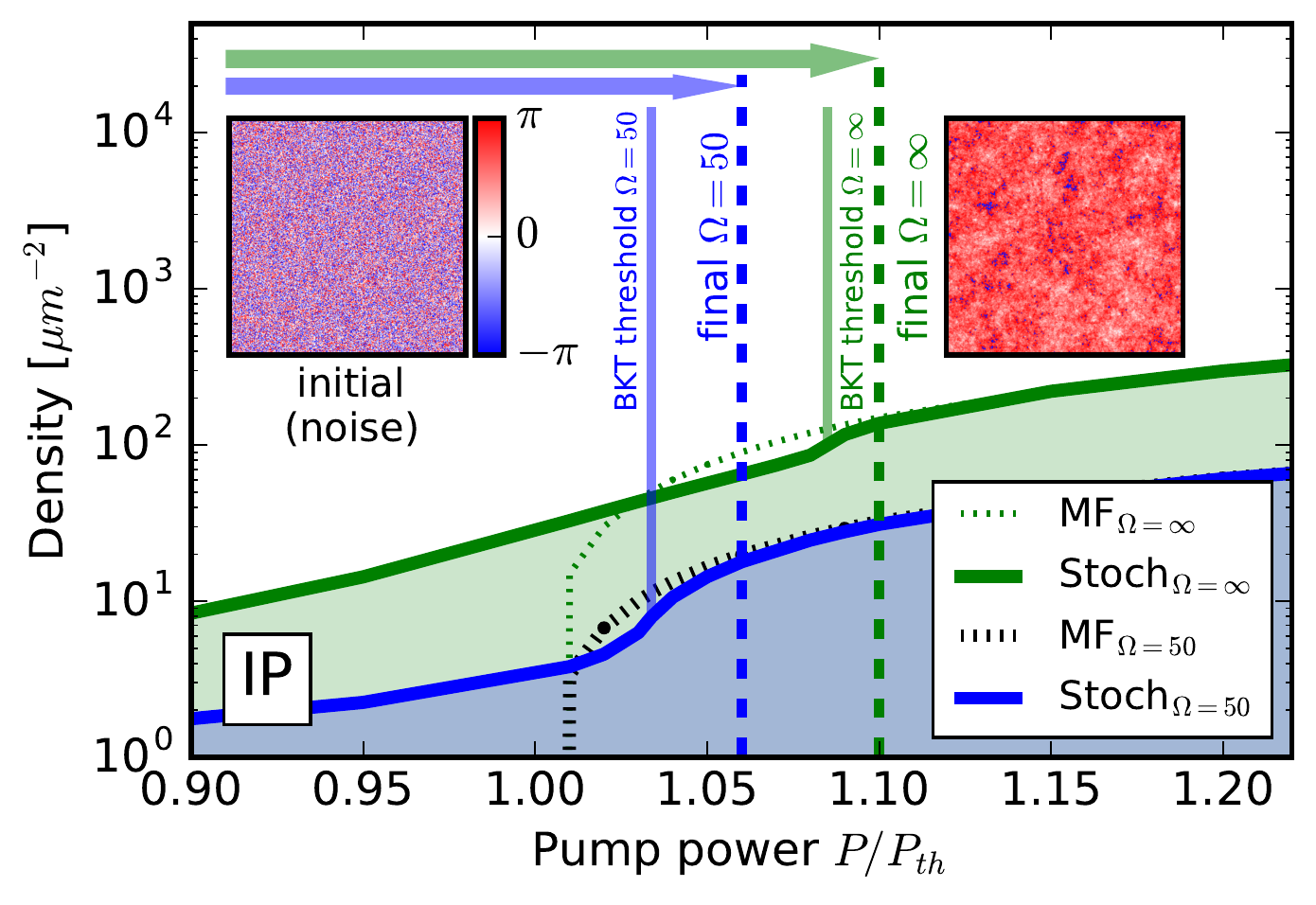}
\caption{{\textbf{BKT transition with parametric and incoherent pumping.}}
Top panel: noise-averaged density from stochastic equations of the
signal (blue), idler (red) and pump (green) as a function
of  pump power $f_p$ for {parametrically} driven
polaritons across OPO.
Bottom panel: mean-field (dotted lines) and noise-averaged (solid
lines) densities for frequency-independent (green) and
frequency-dependent (blue) incoherent pumping.  %
In both panels arrows indicate the infinitely rapid quench protocol across
the critical region and pump powers are scaled to their corresponding mean-field threshold values.
The insets show typical snapshots of the real space phase profile for the 
initial and late-time states.
}
\label{fig:IP_OPO_spectrum}
\end{figure}

Since we are interested in the universal properties of
driven-dissipative systems,  we also consider the alternative typical
set-up of incoherent driving.  Under the assumption that the high-energy
reservoir follows adiabatically the condensate evolution and that the exciton and photon are locked into a single lower-polariton branch, the
equation reads ($\hbar=1$) \red{\cite{chiocchetta2013non}}
\begin{multline}
\hspace{-4mm} id \psi_{LP} = dt\bigg[ - \frac{
\nabla^2}{2 m_{LP} } +
g_{LP}|{\psi_{LP}}|^2_{-} + \frac{i }{2}
\bigg( \frac{P}{1+\frac{|{\psi_{LP}}|^2_{-}}{n_{s}}} -
\gamma_{LP} \bigg)    \\
+\frac{1 }{2}\frac{P}{\Omega} \frac{\partial}{\partial t} \bigg]
\psi_{LP} + \sqrt{\frac{P+\gamma_{LP}}{4}} dW_{LP}
\label{SGPE_IP}
\end{multline}
with ${\psi}_{LP} = {\psi}_{LP}(\textbf{r},t)$ and the lower-polariton
field density reads (after subtracting the Wigner commutator contribution) $|{\psi_{LP}}|^2_{-} \equiv
\left(\left|{\psi}_{LP} \right|^2 - {1}/{dV} \right)$. The Wiener noise $dW_{LP}$ has zero-mean and  fulfils $\langle
dW_{LP}^{*}(\vect{r},t) dW_{LP} (\vect{r}',t) \rangle
=2 \delta_{\vect{r},\vect{r}'}dt$. 
$P$ defines the incoherent saturable and
homogeneous driving strength and $n_s$ the saturation density, and we restrict ourselves to values $\left| \psi_{LP}\right|^2_{-} \ll n_{s}$. We use
typical experimental parameters \red{\cite{nitsche2014algebraic}}:
$\gamma_{LP} = 1/\tau_{LP}$ with  the polariton
lifetime $\tau_{LP} = 4.5 ps$, polariton mass {$m_{LP} = 6.2 \ 10^{-5} \ m_e$},
polariton-polariton interaction strength $g_{LP} = 6.82 \ 10^{-3} \ meV \mu m^2 $.  
To improve the physical relevance of the model when approaching the
critical region, following
\red{\cite{wouters2010superfluidity,chiocchetta2013non}} we implement
frequency-selective pumping mechanism so that relaxation to low-energy
modes is favoured over energies higher than the cut-off frequency
$\omega_{cut} \simeq {\Omega}/{\left(1+ {\left(|\psi_{LP}|_{-}^2 \right)}/{n_s}\right)}$ which are
now suppressed.  We use here $P/P_{th} = 1.06 $, where $P_{th} = \gamma_{LP}$ is the mean field critical pump, $\Omega = 50 \gamma_{LP} = 11.09 ps^{-1}$ and $n_s = 500 \mu m^{-2}$ (labelled as IP$_{\Omega = 50}$), 
for the frequency-dependent pumping, and $P/P_{th} = 1.1 $, $\Omega = \infty$ and $n_s= 1500 \mu m^{-2}$ for the frequency-independent driving (labelled
IP$_{\Omega = \infty}$).
We solve Eq. (\ref{SGPE_IP}) in a square lattice of $301^2$ points, lengths $L_x =L_y= 295.11 \mu m$ and periodic boundary condition and average over a sufficiently-large number (400) of realisations in all schemes.
The healing lengths $\xi =
1/\sqrt{2m_{C,LP} \ g_{C,LP} \ \left|{\psi_{C,LP}}\right|_{-}^2}$ at the end of the evolution are respectively
$\xi^{IP}_{\Omega=50} \simeq 2.2 \mu m$, $\xi^{IP}_{\Omega=\infty} \simeq 0.8 \mu m$ and  $\xi^{OPO} \simeq 1.84 \mu m$, with the condition $L(t) \gg \xi$ fulfilled in all cases \red{\cite{SM}}.

In both models a non-equilibrium steady state with finite particle
density $|\psi_{X,C,LP}|^2$ is established once the pumping strength
overcomes the cavity losses.  Also by tuning the
strength of the pump power both the OPO and the IP system (with/without frequency-dependent pumping mechanism)  undergo a BKT-type of phase transition between a disordered phase with
exponential and an ordered phase with power-law decay of spatial
coherence, governed by the binding/unbinding of vortex-antivortex
pairs (see Fig. \ref{fig:IP_OPO_spectrum}). For the parametric pumping,
the BKT transition in the steady-state is analysed in detail in
\red{\cite{dagvadorj2015nonequilibrium}}. A similar transition takes place for
IP system (Fig. \ref{fig:IP_OPO_spectrum} bottom). It is
worth noting that in this case, stronger fluctuations at higher modes
($\Omega \rightarrow \infty$) and smaller saturation density ($n_s
\rightarrow 0$) lead to a larger shift of the pump power of the BKT
transition with respect to the mean-field onset of
macroscopic population growth.
\begin{figure}
\includegraphics[width=\linewidth]{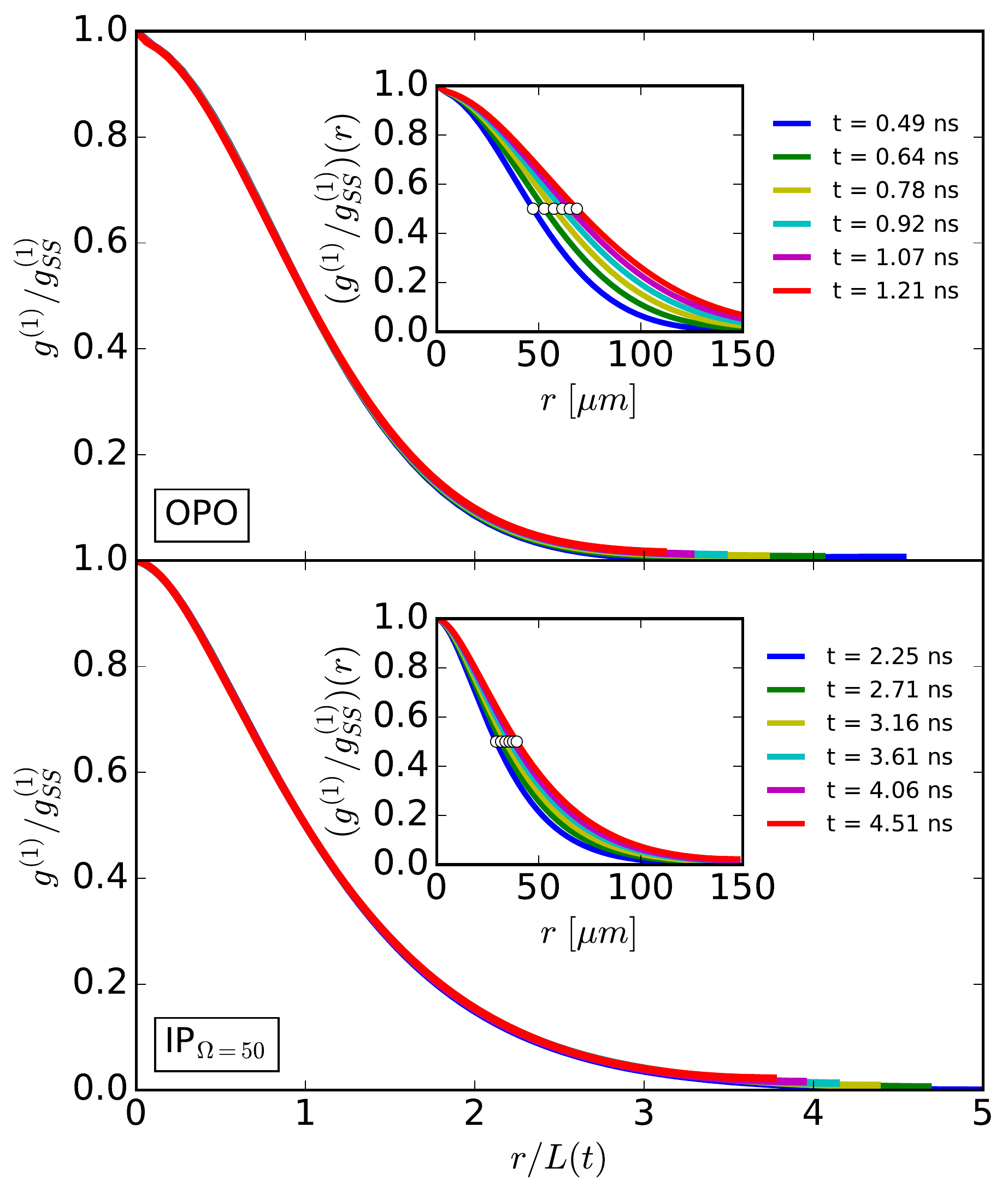}
\caption{ \textbf{Scaling of the two-point correlation function.}
First-order correlation function $g^{(1)}$ (normalised by the corresponding steady state correlator $g^{(1)}_{SS}$) for the
parametrically-pumped (top) and incoherently-pumped polaritons with (bottom) frequency-dependent pumping as a function of
the rescaled distance $r/L(t)$ at different times during
the phase-ordering process.
The apparent collapse of the curves
confirms the dynamic scaling hypothesis.
{Insets show  $g^{(1)} / g^{(1)}_{SS}$  at different times, from which the  characteristic length-scale $L(t)$ is obtained by considering 
$(g^{(1)}/g^{(1)}_{SS})(L(t),t)= 0.5$ (white dots). } 
}
\label{fig:g1_collapse}
\end{figure}

\paragraph*{Universal dynamical scaling.}

\label{sec:universal_scaling_results}

We now study an infinitely-rapid quench across a critical point.  For
the {parametrically} pumped case   (see
Fig. \ref{fig:IP_OPO_spectrum} top), we quench through the upper critical
threshold $f^{\textrm{up}}_p$. The system is prepared in the
steady-state of a deeply disordered phase at a given pump power
$f^i_p > f^{\textrm{up}}_p $ where bound and unbound 
vortices proliferate (see the inset), and is
instantaneously quenched to a deep quasi-ordered regime by adjusting
the external drive to $f^f_p$, with
$f^{f}_p <  f^{up}_p$.
For the incoherently pumped case (see
Fig. \ref{fig:IP_OPO_spectrum} bottom), we 
quench from the random initial configuration deep in the disordered phase (i.e. steady-state for $P^i = 0$), to a quasi-ordered regime by setting the external drive to $P^f > P_{th}$ at $t=0$ and letting the system evolve.
We explore the dynamical scaling properties of the system during the
phase ordering by considering the 
first order two-point
correlation function
\red{\cite{dagvadorj2015nonequilibrium}}: \beq g^{(1)} (\vect{r},t) =
\Frac{\langle \psi^* (\vect{r} + \vect{u},t) \psi^{} (\vect{u},t)
\rangle -{\delta_{\vect{r} + \vect{u},\vect{u}}}/{2 dV}}{\sqrt{
\langle \left| \psi(\vect{r} + \vect{u},t)\right|^2_{-}
\rangle \langle \left| \psi(\vect{u},t) \right|^2_{-}
\rangle}}\; ,
\label{eq:corre}
\eeq
where $\langle \dots \rangle$ denotes averaging over both noise
realisations and the auxiliary position $\vect{u}$, and $t$ is the
time after the quench.  For the
parametrically pumped scheme, $\psi$ in Eq.~(\ref{eq:corre}) corresponds to
the signal, which is obtained by filtering the cavity-photon field 
$\psi_C$  from Eq. \eqref{eq:wigner} around the signal momentum
$\textbf{k}_s$. For incoherently pumped system, $\psi$ is the
polariton field $\psi_{LP}$.

In the phase ordering kinetics of the planar
XY-model in 2-dimensions
\red{\cite{jelic2011quench,bray2000breakdown,rutenberg1995phase}},
the non-steady state two-point correlator
\eqref{eq:corre} fulfils the dynamical scaling
form:
{$ g^{(1)} ({r},t) \sim g^{(1)}_{SS}({r},t) \cdot F\left( {r}/{L(t)}  \right)
$}, {with the steady state correlation 
function decaying algebraically at long distances, as $g^{(1)}_{SS}(r) \sim r^{-\alpha}$ .}
The scaling function $F$ tends to $1$ when $ r \ll L(t)$, indicating
that the critical correlations have been established at distances much
smaller than $L(t)$ at time $t$, which defines the \emph{characteristic
length-scale} of the system $L(t)$. Since our system is highly
non-equilibrium it is far from obvious whether similar scaling
behaviour holds here in the presence of strong drive and
dissipation. {Indeed, we obtain a perfect collapse when plotting the two
point correlation function {$g^{(1)}/ g^{(1)}_{SS}$} as a function of the rescaled
length $r/L(t)$ at different times of the late dynamics for both
driving schemes (see Fig. \ref{fig:g1_collapse}).  For consistency,
we extract the length-scale $L(t)$ when {$(g^{(1)}/g^{(1)}_{SS})(L(t),t)= 0.5$}  (white dots in insets
of Fig.~\ref{fig:g1_collapse}), with the independence of our conclusions on the intersection value verified in the Supplementary Material \red{\cite{SM}}.}

Since our polariton system fulfils the scaling hypothesis, as revealed
by the collapse of the two-point correlation function, we can access
the universal dynamical critical exponent $z$ of our
driven-dissipative system by analysing the growth of the
characteristic length L(t) and the decay of the number of topological
defects (vortices) at late times after a sudden quench.  Note, that an
equilibrium analogue of the phase degree of freedom for polariton
system is the planar XY-model, where free vortices and bound vortex-antivortex pairs exist even in the steady-state below and above the
BKT phase transition respectively. The existence of the steady-state
vortices plays a fundamental role in the phase ordering process and
introduces the characteristic logarithmic correction into the
late-time dynamics both of $L(t)$ and number of vortices $n_v$
following a sudden quench
\red{\cite{yurke1993coarsening,jelic2011quench}}, such that $L (t) \sim
((t/t_0)/\log(t/t_0))^{1/z}$ and $n_v(t) \sim ((t/t_0)/\log(t/t_0))^{-(2/z)}$ where $t_0$ is a nonuniversal microscopic system timescale (taken here as $t_0 = 1 ps$ \red{\cite{SM}}). The two
relations follow from the fact that $n_v(t) \sim 1/L(t)^2$ when
there is a \emph{unique} length-scale in the system, which is true
at late times in the dynamics \red{\cite{SM}}. 

\begin{figure}
\includegraphics[width=1\linewidth]{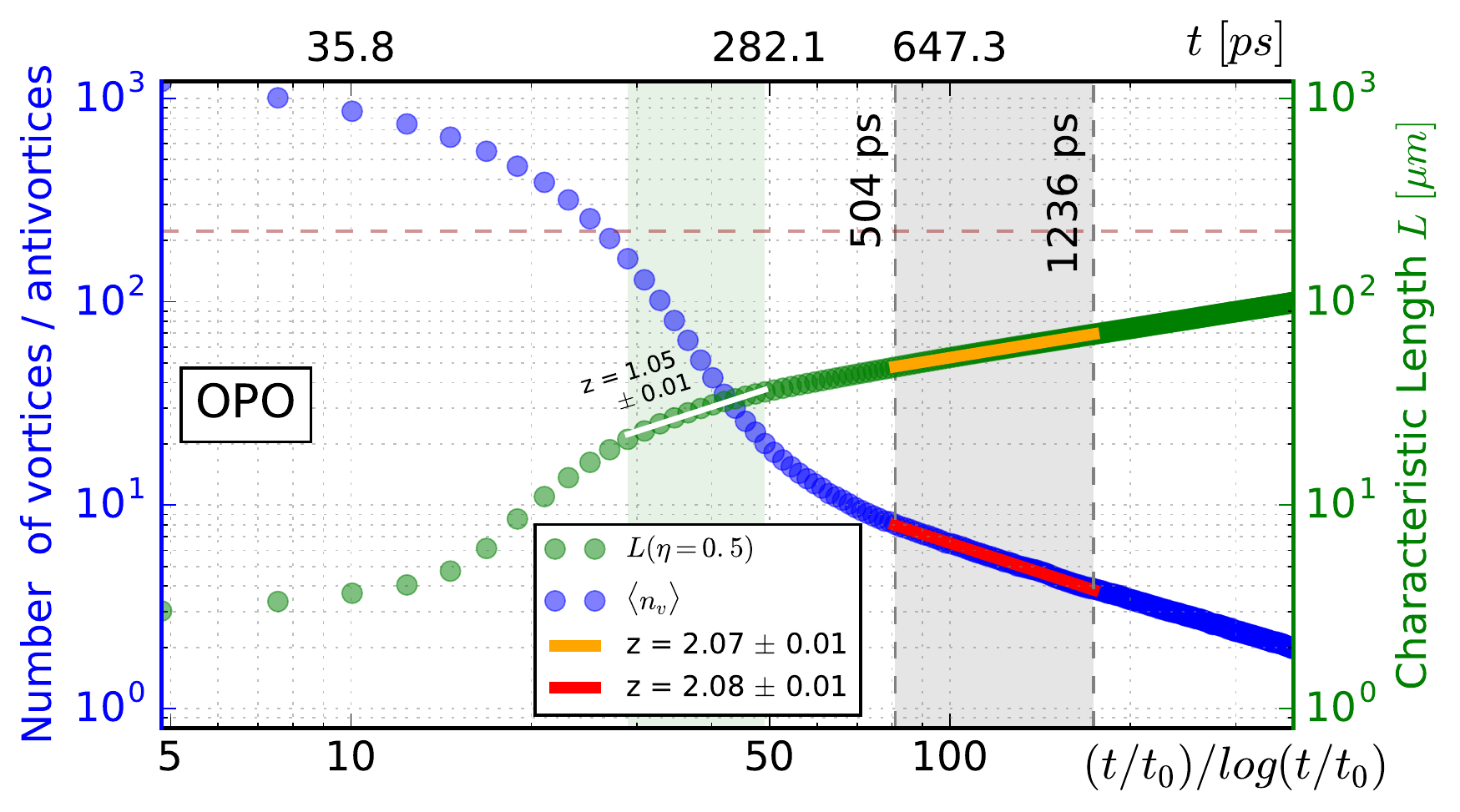}
\includegraphics[width=1\linewidth]{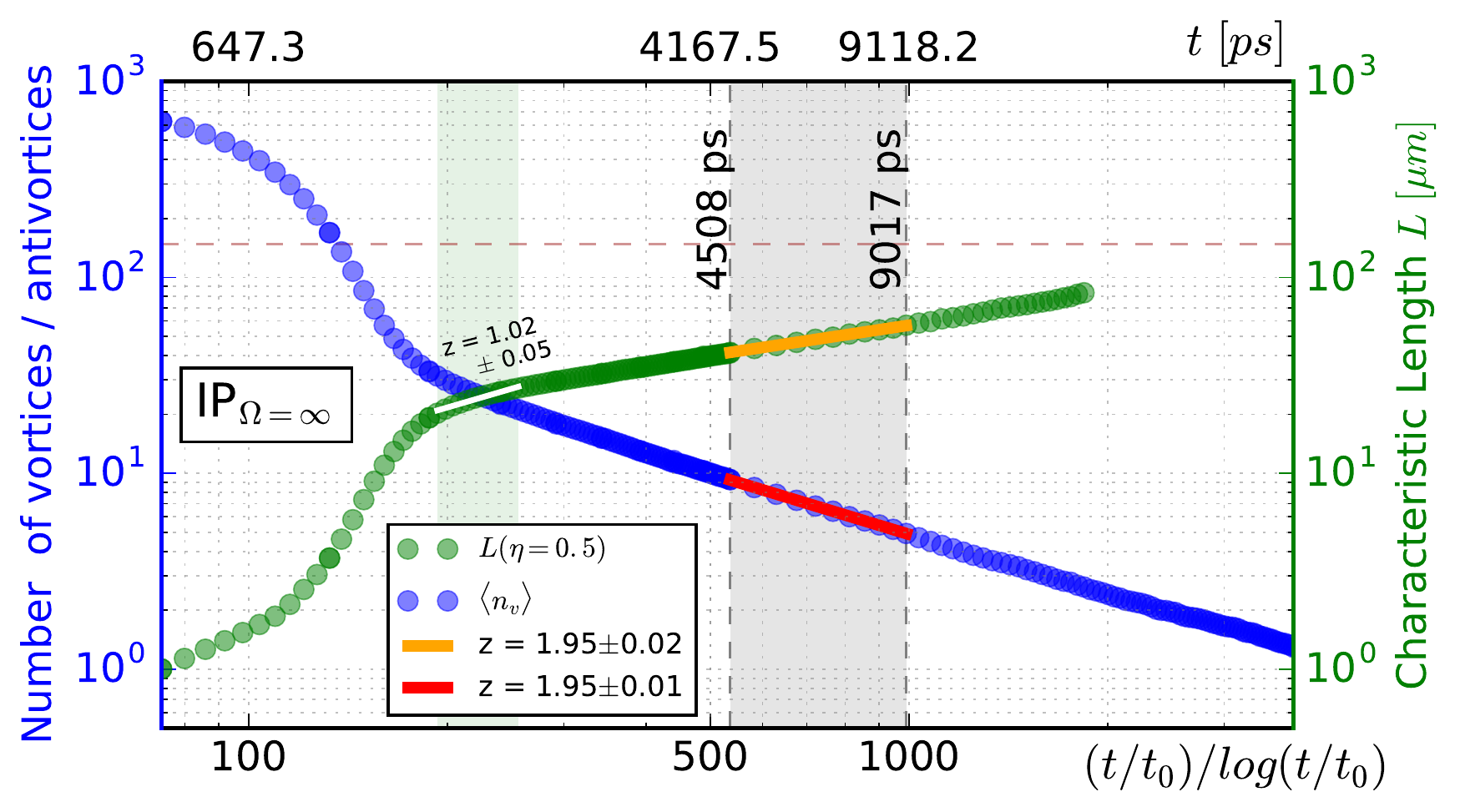}
\includegraphics[width=1\linewidth]{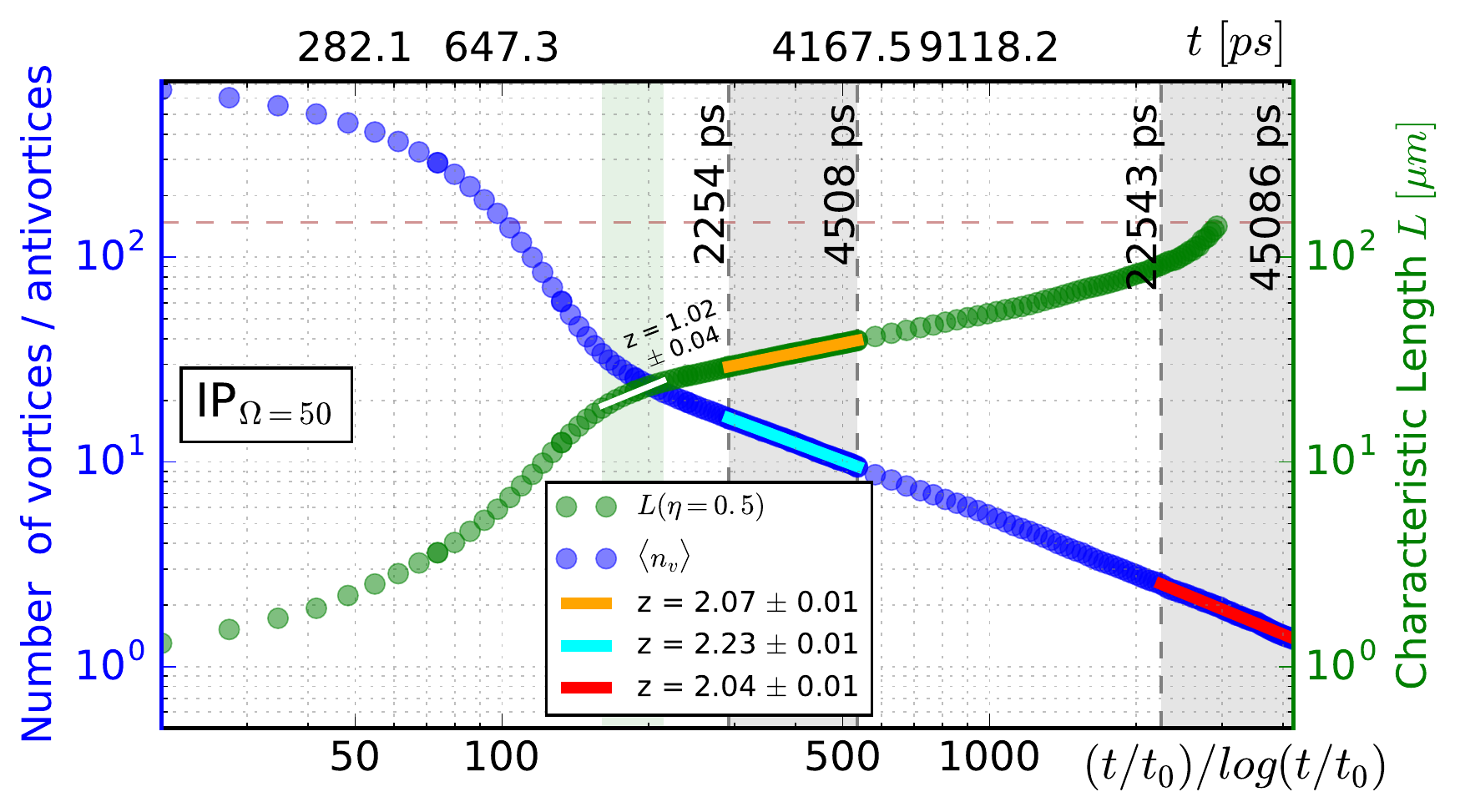}
\caption{\textbf{Topological defects and  $L(t)$ during
phase ordering}. Density of vortices (blue) and the
characteristic length-scale $L(t)$ (green) as a function of
time after an infinite rapid quench for parametric (top panel), 
frequency-independent (central panel) and frequency-dependent
(bottom panel) incoherently-pumped polaritons. The size of the numerical grid is marked by the horizontal (brown) dashed lines. 
Late-time dynamics show characteristic diffusive
behaviour described by logarithmic corrections $(t/t_0)/\log(t/t_0)$ to the
dominant power-law scaling  due to the
presence of the topological defects. For all configurations, we
obtain a non-equilibrium dynamical critical exponent $z\approx 2$ within the grey-shaded regions.
}
\label{fig:vortices_vs_time}
\end{figure}

To demonstrate the emergence of this universal scaling from our numerical data, Fig. \ref{fig:vortices_vs_time} plots the characteristic length
$L(t)$ and number of vortices $n_v(t)$ for the three different pumping
schemes considered. 
OPO (top) and IP$_{\Omega = \infty}$ (centre)
cases are qualitatively identical in the sense that $g^{(1)}$ collapses in the same time window as that in which the respective vortex dynamics reaches the converged $z\approx2 $ value.  
We are careful to fit $L(t)$ and $n_v$ late enough in 
the quench so the dynamics becomes universal (indeed both reveal $z \approx 2$) but before size effects, power-law correlations or very small vortex number affect our analysis.
However, in the case of IP$_{\Omega = 50}$ (bottom),  strong damping of collective fluctuations, introduced by the explicitly
frequency-dependent nature of the pump, is responsible for the 
collective modes to reach $z\approx 2$ at a much earlier time than 
the topological modes.
Nevertheless, both channels show the dynamical critical exponents to be $z\approx 2$ in their appropriate late time windows.  
A more detailed discussion about the fitting criteria can be found in Supplementary Material \red{\cite{SM}}.

We stress that the sufficiently late-time analysis is essential to
allow all channels to equilibrate properly, and fulfil the scaling
hypothesis, whereas fitting early in the phase-ordering process, and before the dynamics becomes universal (light green regions in Fig. \ref{fig:vortices_vs_time}), can lead to the incorrect conclusion of $z \approx 1$. 
Note that for the experimentally realistic parameters considered in our 
simulations, the phase-ordering takes place on timescales similar to the 
polariton population growth \red{\cite{SM}}, which is the 
case for which Ref.~\red{\cite{kulczykowski2017phase}} predicted $z\approx 1$.

\paragraph*{Summary and outlook.}

We have extended the study of universal critical properties,
specifically the dynamical critical exponent, of strongly
driven-dissipative two-dimensional quantum systems by considering an
infinitely rapid quench from the disordered to a quasi-ordered
phase. Our work reveals clearly that such universal properties of the system only emerge for appropriately late time dynamics of both the topological defects and smooth phase fluctuations, whereas the early dynamics are nonuniversal and can give misleading information
about the value of the critical exponents.

Importantly, for all pumping configurations we find a value $z\approx 
2$ for the dynamical critical exponent, different from the one of 
conservative Bose systems \red{\cite{williamson2016universal,kudo2013magnetic,damle1996phase,hofmann2014coarsening}}.
On the other hand, our results indicate that under realistic polariton pumping 
schemes and experimental parameters, the relevant scaling is analogous 
to the one predicted for the planar XY-model \red{\cite{jelic2011quench,bray2000breakdown,rutenberg1995phase}} and, therefore, the non-equilibrium physics brought by the KPZ nonlinearity does not show in the phase ordering of realistic size systems. 

\paragraph*{Acknowledgements.}

We would like to thank George W. Stagg and Kean Loon Lee for computational assistance and Michal Matuszewski and Leticia Cugliandolo for fruitful discussions. 
We acknowledge financial support from EPRSC: Grants Nos. EP/I028900/2 and EP/K003623/2 (AZ, GD and MHS) and EP/L504828/1 (PC and NPP for DTA support).
This work was supported by the EU-FET Proactive grant AQuS, Project No. 640800, and by the Autonomous Province of Trento, partially through the project ``On silicon chip quantum optics for quantum computing and secure communications (SiQuro)."



\pagebreak

\widetext
\begin{center}
	\textbf{\large Supplementary Material for: 			Dynamical critical exponents in driven-dissipative quantum systems}
\end{center}

\vspace{8mm}

\setcounter{equation}{0}
\setcounter{figure}{0}
\setcounter{table}{0}
\setcounter{page}{1}
\renewcommand{\theequation}{S\arabic{equation}}
\renewcommand{\thefigure}{S\arabic{figure}}
\renewcommand{\bibnumfmt}[1]{[S#1]}
\renewcommand{\citenumfont}[1]{S#1}

%
%
%
%
%
%
%
%
%
%
%
%
%


In this Supplementary Material we present technical details related to our numerical procedure and analysis, offering conclusive proof that $z \approx 2$  with logarithmic corrections is the correct dynamical exponent for exciton-polariton systems across all experimentally-relevant regimes.
After explicitly demonstrating numerical convergence, and the different physical regimes probed in terms of the interplay of density-growth and vortex-decay dynamics, we demonstrate the clear emergence of $z\approx 2$ in the presence of logarithmic corrections at sufficiently late evolution times;
we also show the direct relation between vortex number and obtained characteristic length scale, demonstrate the effective independence of our conclusions on the choice of intersection point for the correlation function collapse and show that careful consideration of the conventionally-ignored intrinsic nonuniversal time-scale $t_0$ does not affect our findings.

\vspace{8mm}

\twocolumngrid

\section{Numerical Methods and Convergence}
\label{app:convergence}

\paragraph{First order correlation function.} 
For both parametric and incoherently pumped case, the first order correlation function $g^{(1)}(r)$ is calculated by means of Eq.~(\ref{eq:corre}) in the main text. 

For the incoherently pumped case, we first compute the two-point correlation function for each vertical and horizontal array of a single realisation of the wave function $\psi$. The resulting $N \times N $ outcomes are then averaged and normalized before additionally averaging over $N_p$ realizations, and computing corresponding error bars.
For the parametric pump case, we filter the full emission $\psi_C$, in such a way as to omit contributions with momentum outside a set radius about the signal states, in momentum space with a sharp step-like rectangular filter; for more details see \red{\cite{Adagvadorj2015nonequilibrium}}.  
The noise-averaged autocorrelation function of the filtered field $\psi_s$ is computed efficiently in momentum space.

\paragraph{Vortex Number Calculations.}
The number of vortices is evaluated from the phase gradients around closed paths of each grid point. 
We extract the phase, $\phi$, and gradient of the phase, $\nabla \phi = v$, using finite differences.
A grid point is identified as having a vortex when the circulation $\Gamma = \int_{C} v \ d\bf{r} \gtrsim 2 \pi$. 

For the incoherently pumped system, a Gaussian (low-pass) filter is first applied to the wave function, in order to remove all the high frequency noise components. This removes all noise with 
wavelength (in pixels) smaller than, or of the order of, the standard deviation of the filter's Gaussian kernel.

\begin{figure}
	\includegraphics[width=1\columnwidth]{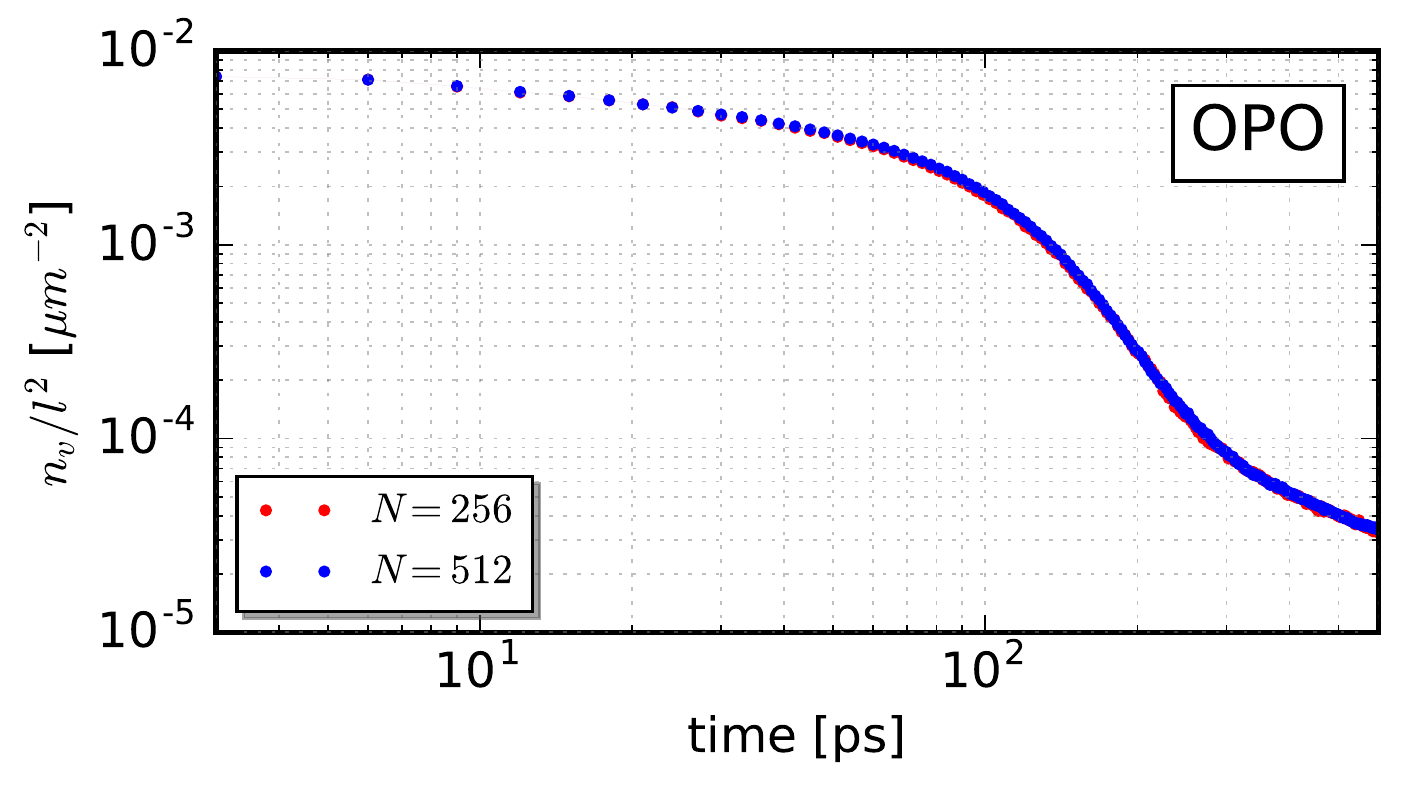}
	\includegraphics[width=1\columnwidth]{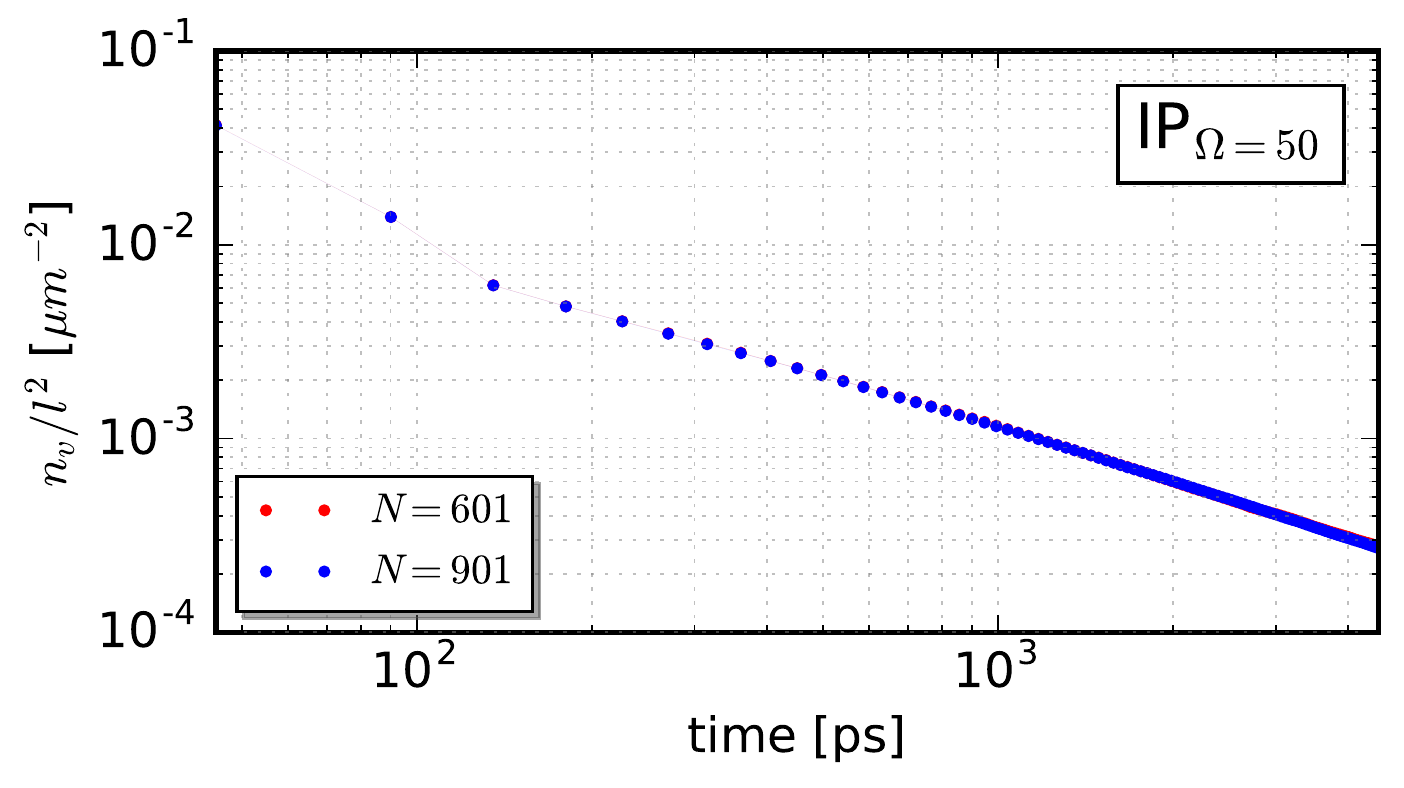}
	\includegraphics[width=1\columnwidth]{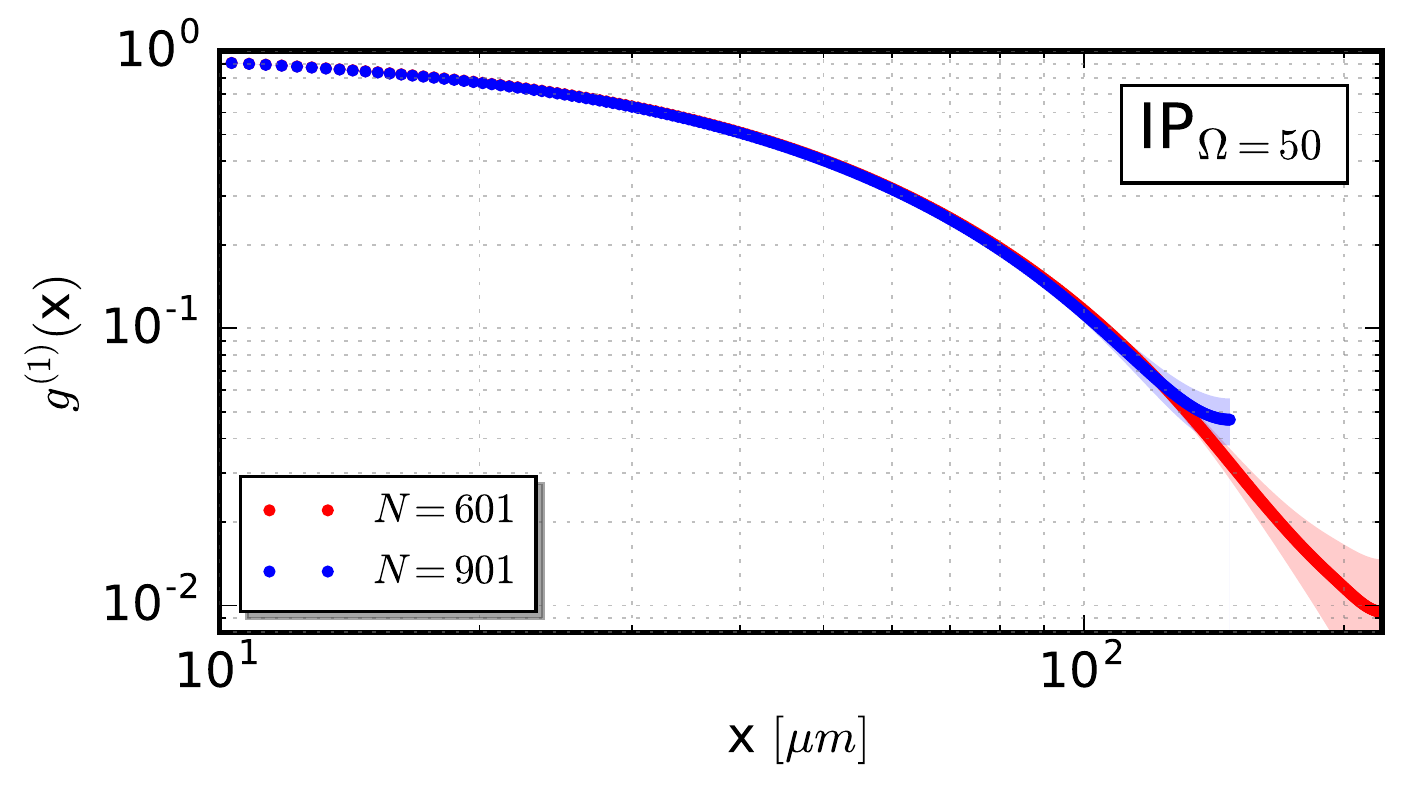}
	\caption{
		\textbf{Convergence in lattice size.} 
		Top panel: convergence in time of vortex density for OPO system with $f_p / f^{up}_p  = 0.97$, for two different box of size $l = 222.208 \ \mu m$ and $l = 444.42 \ \mu m$  with respectively $N=256$ and $N=512$ number of points and grid spacing $a = 0.87 \ \mu m$. 
		Central panel: convergence in time of vortex density for IP system with $P / P_{th} = 1.5$, $\Omega=11.09 \ ps^{-1}$ and $n_s = 500 \ \mu m^{-2}$, for two different box of size $l=295.11 \mu m$, $l= 444.42 \ \mu m$  with respectively $N=601$ and $N=901$ number of points. 
		Bottom panel: spatial convergence of first order correlation function for IP system before it enters the power-law stage at $t=4.5 \ ns$ for the two different boxes.
	}
	\label{fig:convergence_size}
\end{figure}

\paragraph{Convergence in lattice size.} 
Fig.~\ref{fig:convergence_size} provides conclusive evidence of the independence of our results on lattice size for both OPO and IP regimes, based on which the analysis in our main paper is conducted for grid spacings $a = 0.87 \mu m$ (OPO), $a = 0.98 \mu m$ (IP) and grid sizes $444.42 \ \mu m$ (OPO) and $l=295.11 \mu m$ (IP). Specifically, the top (middle) panel show a comparison of the time-evolution of the vortex density for the OPO (IP) case upon increasing the total grid size, while keeping discretisation fixed.
Fig.~\ref{fig:convergence_size} (bottom) shows corresponding IP results (for $\Omega=50 \gamma_{LP}$) for the correlation function, with the coloured bands indicating the error bars; this demonstrates that our numerics is sufficient to avoid the correlation function being affected by boundary effects even as the system approaches the critical region.

\begin{figure}
	\includegraphics[width=1\columnwidth]{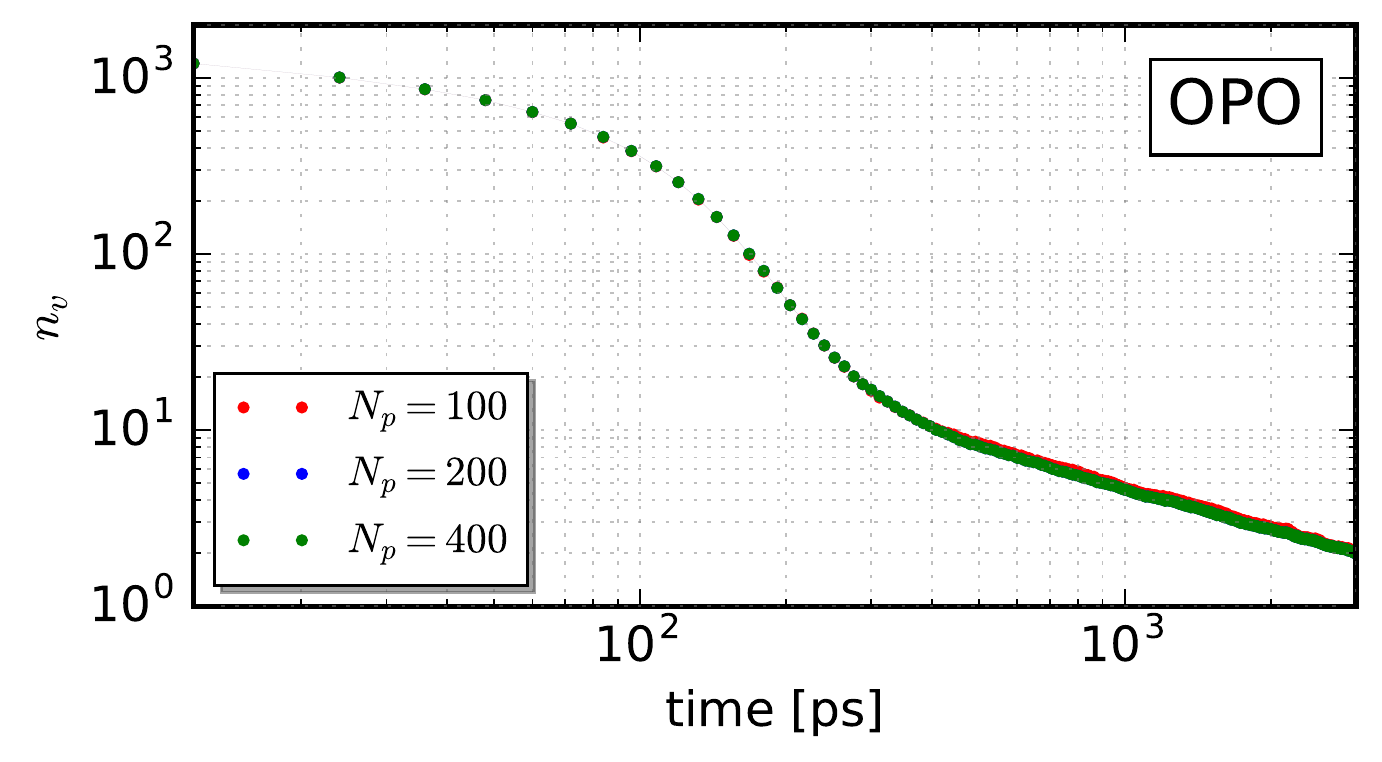}
	\includegraphics[width=1\columnwidth]{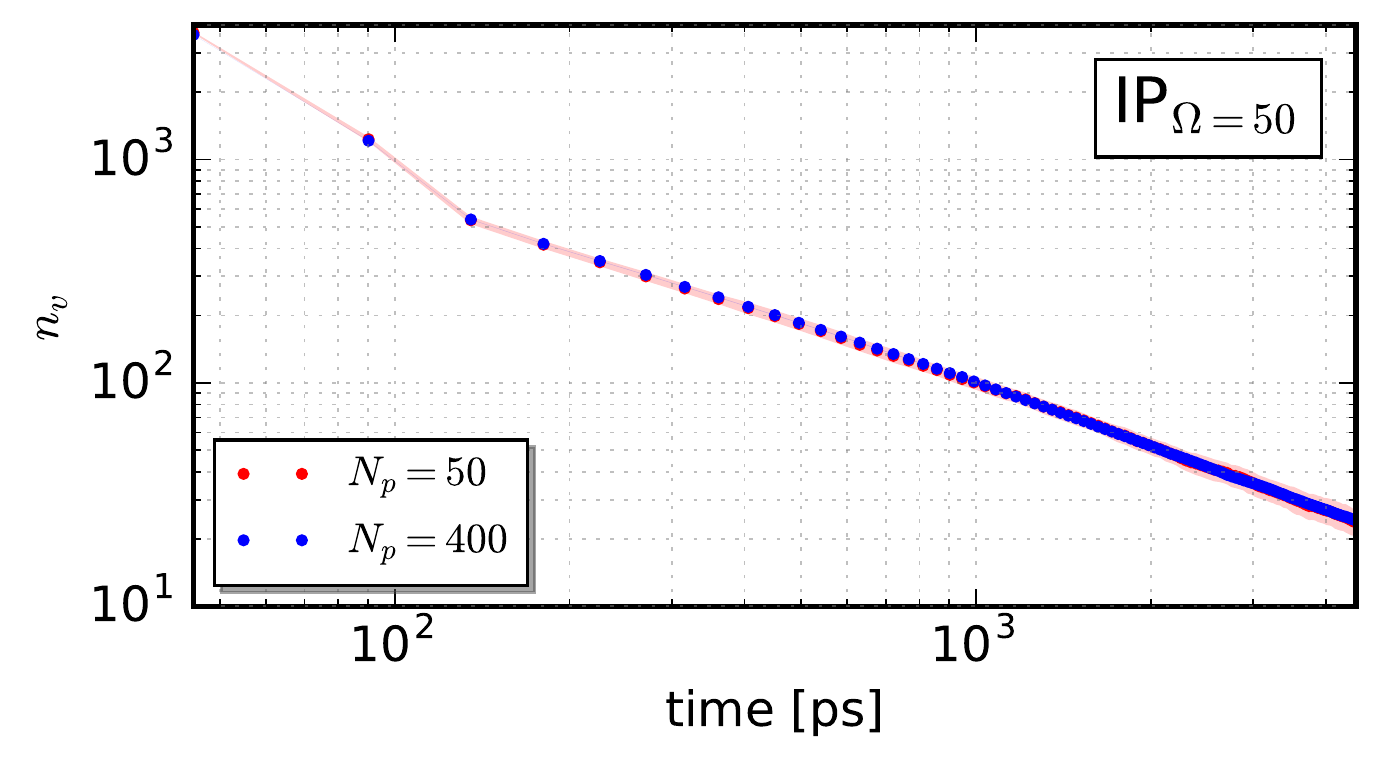}
	\includegraphics[width=1\columnwidth]{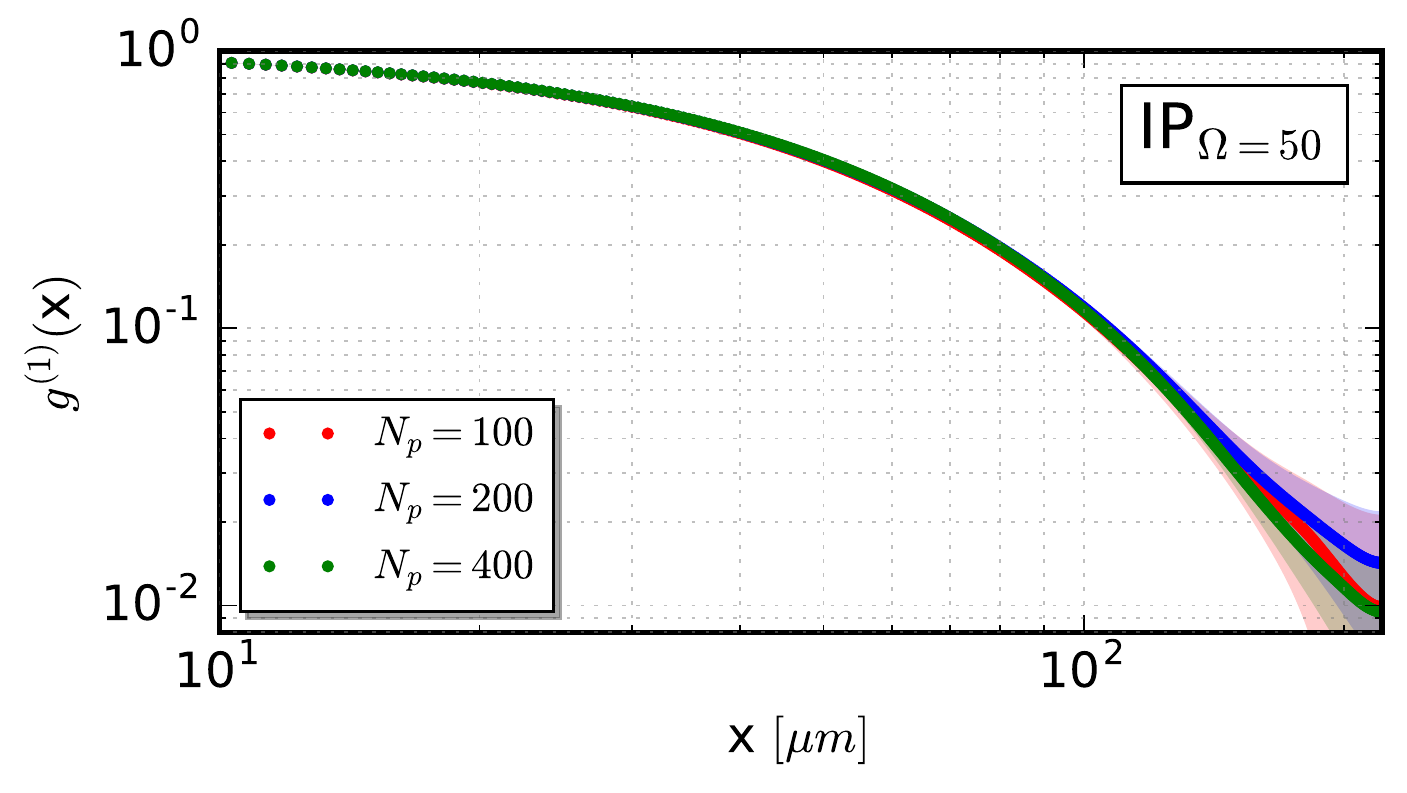}
	\caption{
		\textbf{Convergence in number of stochastic paths.}
		Top panel: convergence in time of number of vortices for OPO system averaged over $N_p= 100, 200, 400$ stochastic paths.
		Central panel: convergence in time of number of vortices for IP system averaged over $N_p= 50, 400$ noise realizations. 
		Bottom panel: spatial convergence of first order correlation function for IP system before it enters the power-law stage at $t=4.5 \ ns$ over $N_p= 100, 200, 400$ stochastic paths. Parameters as in Fig.~\ref{fig:convergence_size}.
	}
	\label{fig:convergence_realizations}
\end{figure}

\paragraph{Convergence in stochastic numerical realisations.} 
Having fixed the grid spacing and size, we next investigate the dependence of results on realisation over different numerical trajectories (done in addition to averaging over the entire grid in single realisations).
Although Fig.~\ref{fig:convergence_realizations} shows that on first inspection a number of $\approx 100$ (OPO) or 50 (IP) realisations may be enough for sufficient convergence in the overall dynamics, we nonetheless stress that 
averaging over a high number of realisations is essential for an accurate determination of the long-time dynamical critical exponent for $L$ and $n_v$. Throughout our work we have therefore used $N_p = 400$ realisations for both OPO and IP cases.

\paragraph{Convergence in computational method (Incoherently pumped system).} 
\begin{figure}
	\includegraphics[width=1\columnwidth]{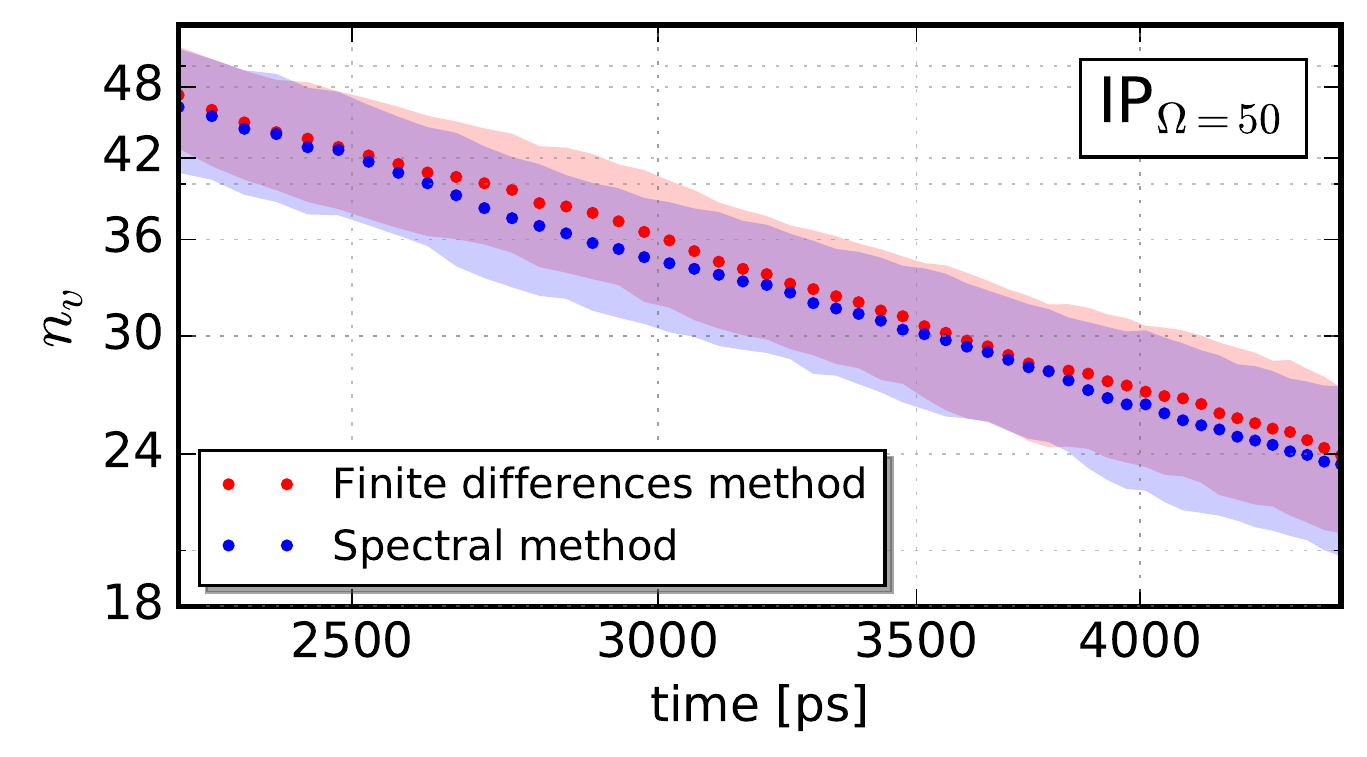}
	\includegraphics[width=1\columnwidth]{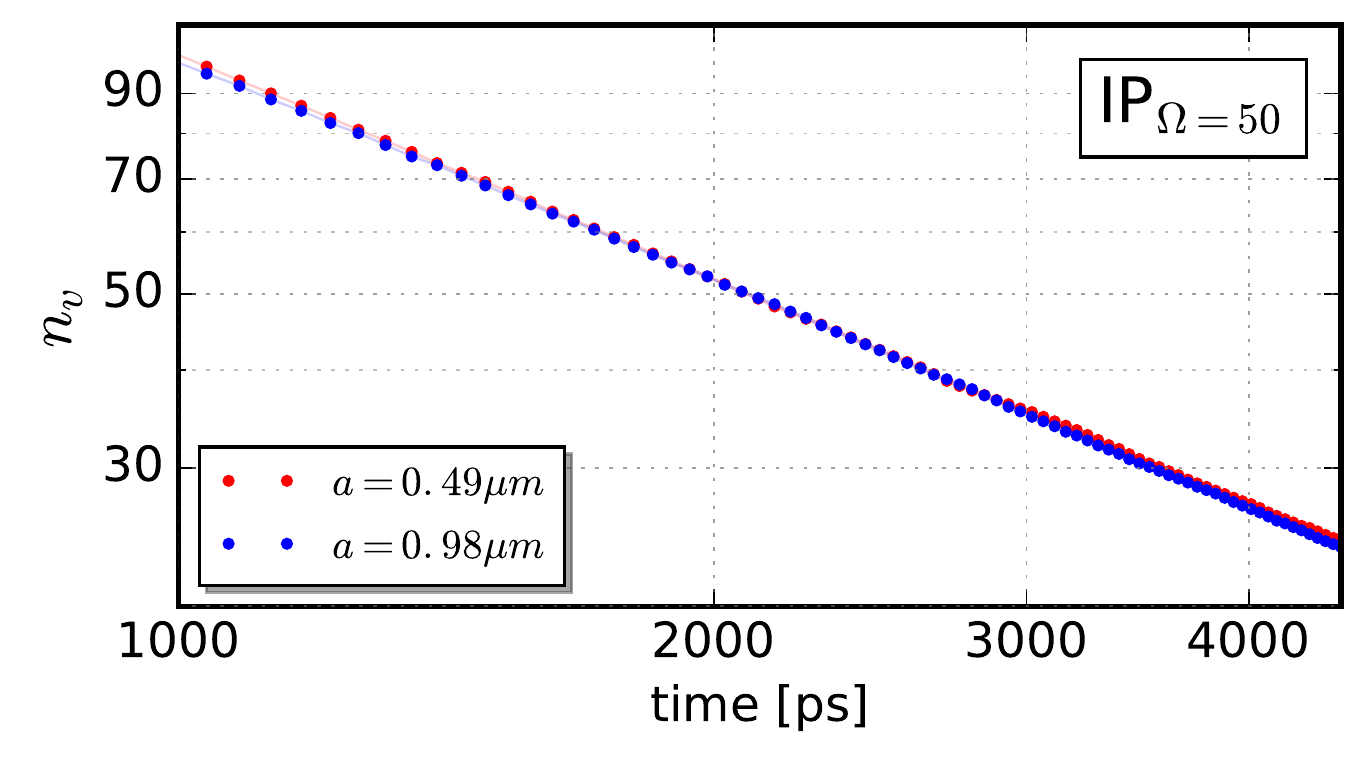}
	\caption{
		\textbf{Convergence in methods and cut-off choice for the incoherently pumped system. } Top panel: evolution of number of vortices for two different computational schemes: spectral methods (red points) and finite differences methods (blue points). Bottom panel: time evolution of number of vortices for two different lattice spacings, i.e. cut-off choice, $a = 0.49 \mu m$ (red points) and $a = 0.98 \mu m$ (blue points).
		Parameters as in Fig.~\ref{fig:convergence_size}.
	}
	\label{fig:convergence_models_cutoff}
\end{figure}

To ensure independence of our numerical results on computational method, we have implemented two distinct schemes for the incoherently-pumped case, based on finite differences method for the Laplacian implemented in Fortran and  based on spectral methods and the publicly-available XMDS2 code \red{\cite{Ajournals/cphysics/DennisHJ13}}. 
Fig.~\ref{fig:convergence_models_cutoff} (top) shows excellent agreement between the two cases even for only 50 numerical realisations. Throughout our work, we have chosen to report results based on the XMDS2 spectral method, as this has exponential convergence, i.e. the error scales as $\propto a^N$ with increasing resolution (number of grid points N) - as opposed to the algebraic scaling $\propto a^2$ of finite difference schemes.

\paragraph{Convergence in cut-off choice (Incoherently pumped system).} 
Within the stochastic Gross-Pitaevskii equation, the grid spacing $a$ sets a maximum momentum cut-off $\propto a^{-1}$ in the numerical representation of the quantum field, restricting the maximum number of evolving modes. It is therefore important to check that our results do not depend on the chosen cut-off value. Fig.~\ref{fig:convergence_models_cutoff} (bottom) demonstrates that for two different grid spacings $a = 0.49 \ \mu m$ and $a = 0.98 \mu m$ (while keeping the total box size fixed to $ l = 295.11 \mu m$ by varying the number of grid points) the evolution of $n_v$ is practically identical.

\section{Vortex number dynamics}
\label{app:dynamics}

A previous study \red{\cite{Akulczykowski2017phase}} of incoherently-pumped  exciton-polariton systems has observed the dynamical critical exponent to be dependent on the quality of the sample, through the polariton lifetime, thus arguing for different types of  non-universal dynamics for the driven-dissipative exciton-polariton system.
Contrary to such a statement, our work demonstrates unequivocally that the universality class of the phase transition in polariton systems of experimentally realistic size falls, as anticipated, within the 2D-XY-model universality class provided that exponents are extracted at the appropriate late times.

\begin{figure}
	\includegraphics[width=1\columnwidth]{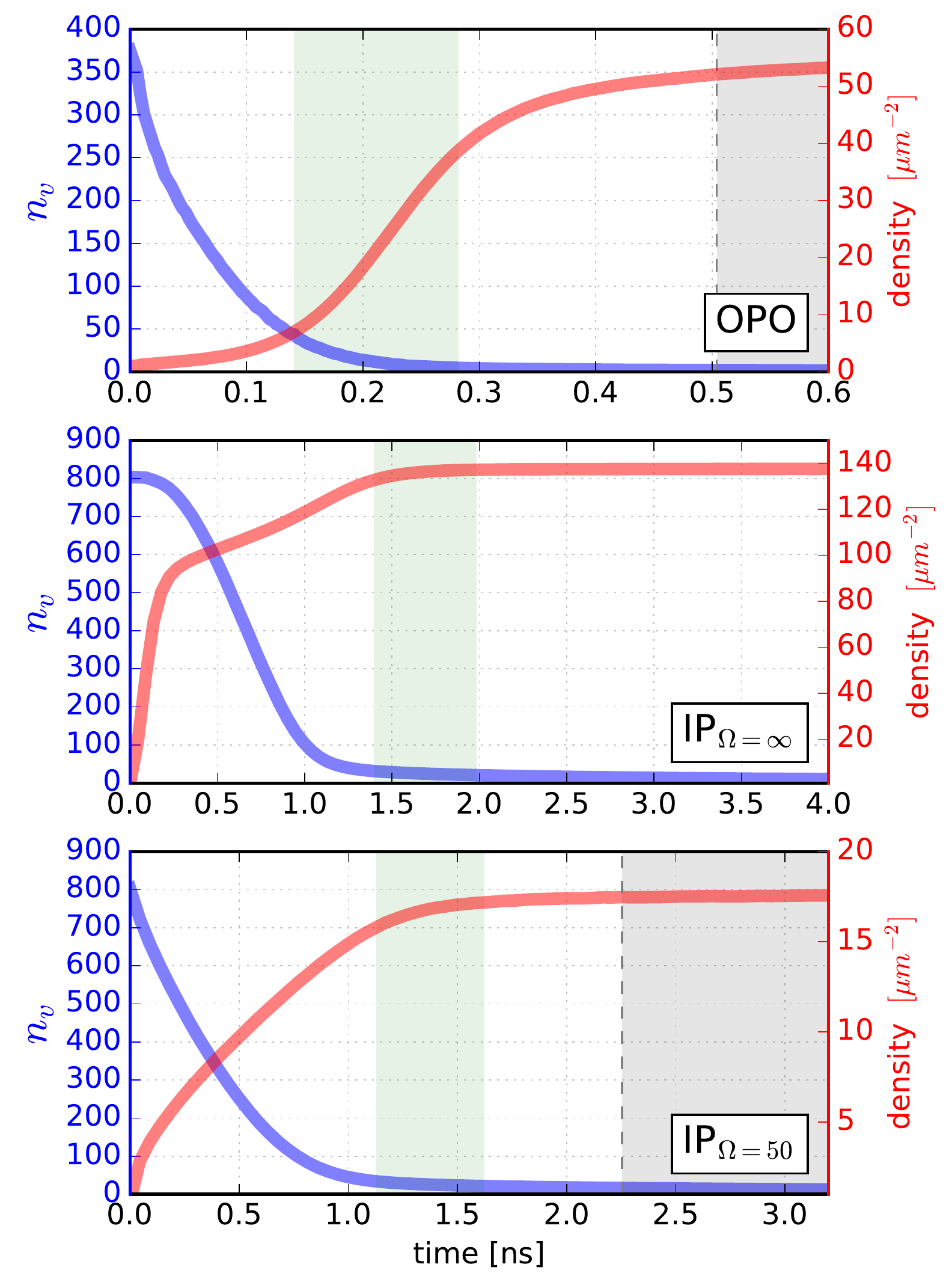}
	\caption{
		\textbf{Evolution of averaged density and number of vortices.}
		Density growth and pair annihilation for coherent (top), frequency-dependent (central) and frequency-independent incoherent pumping systems (bottom). We report regions where vortex dynamics follows Eq.~\eqref{fitz_log} with critical exponent $z \approx 1$ (faint green region) and $z \approx 2$ (grey region) as reported in Fig.~\ref{fig:vortices_vs_time} main paper. Parameters as in main paper. 
	}
	\label{fig:evolution_density_nv}
\end{figure}

\begin{figure*}
	\includegraphics[width=1.8\columnwidth]{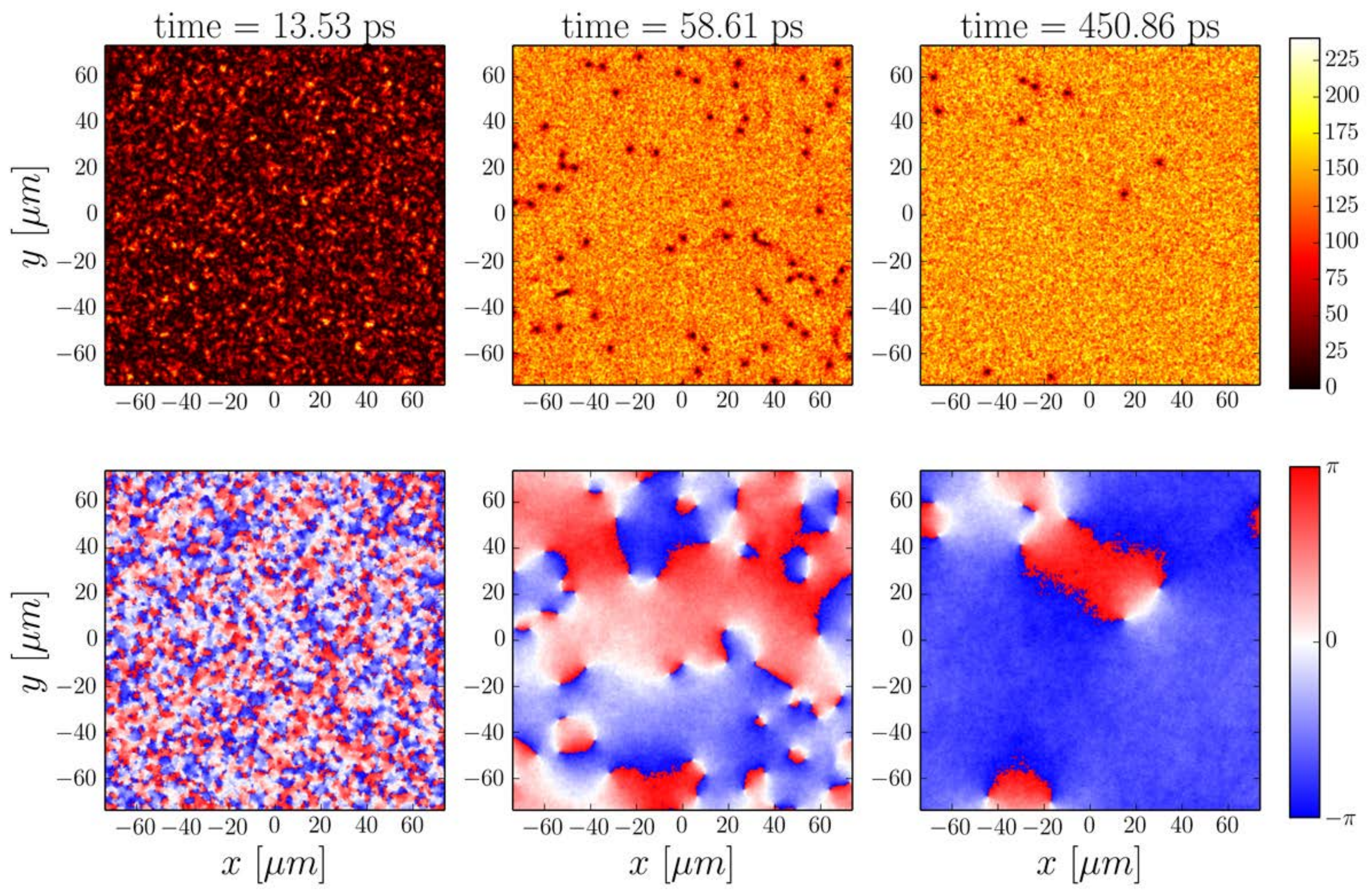}
	\caption{
		\textbf{Phase ordering dynamics of an IP system.}
		Snapshots capturing the phase ordering process for condensate density in $\mu m^{-2}$ (top) and phase (bottom) of an incoherently pumped system. The initially noisy configuration of the system (left) becomes gradually ordered through the annihilation of vortex pairs (centre and right) according to the scaling law of Eq.~(\ref{fitz2_log}). Parameters as in Fig.~\ref{fig:convergence_size}.
	}
	\label{fig:coarsening}
\end{figure*}

Fig.~\ref{fig:evolution_density_nv} displays time evolutions of the averaged particle density and average number of vortices for the OPO (top) and IP (middle/bottom) systems.
During the growth of the density of the degenerate exciton-polariton system, the
vortex pairs decay monotonically in time, eventually reaching the long-time limit where universal phase-ordering kinetics features should set in.

Ref.~\red{\cite{Akulczykowski2017phase}} has effectively argued that the value of the dynamical critical exponent depends on the interplay between time-scales for density saturation and vortex pair annihilation, indicating that in cases where the two processes occur effectively in parallel one would expect the critical exponent $z=1$, characteristic of conservative superfluids \red{\cite{Adamle1996phase,Awilliamson2016universal}}, whereas in the case where the majority of vortex pair annihilation occurs after the density has effectively saturated, one would expect $z=2$, as in the 2D XY model \red{\cite{Abray2000breakdown,Ajelic2011quench}}.

Fig.~\ref{fig:evolution_density_nv} shows the three cases considered within our numerics, based on experimentally-relevant parameters \red{\cite{Akasprzak2006bose,Anitsche2014algebraic}}. Specifically we consider cases in which the two processes occur in parallel (bottom), the density growth is initially much faster than vortex annihilation (middle), and the density grows after most vortex annihilation has taken place (top). As reported in the main text (Fig.~\ref{fig:vortices_vs_time}), when fitting all such cases with 
the applicable scaling law with logarithmic corrections
\begin{equation}
n_v(t) \sim [(t/t_0)/log(t/t_0)]^{-2/{z}},
\label{law_log_nv}
\end{equation}
we find clearly that $z \approx 1$ at early evolution times (highlighted in Fig.~\ref{fig:evolution_density_nv} by the faint green regions), and that the dynamical critical exponent always converges to $z \approx 2$ at later times (grey regions in same plots), as convincingly shown in next section. 

Here $t_0$ is a nonuniversal microscopic system time-scale and consideration of all relevant system time scales (see last section) demonstrates our results to be insensitive to its precise value.

For completeness, Fig.~\ref{fig:coarsening} depicts a characteristic example of density and phase profiles during the coarsening process.

\section{Logarithmic corrections and dynamical exponent}
\label{app:analysys_log_correction}

\begin{figure}
	\includegraphics[width=\columnwidth]{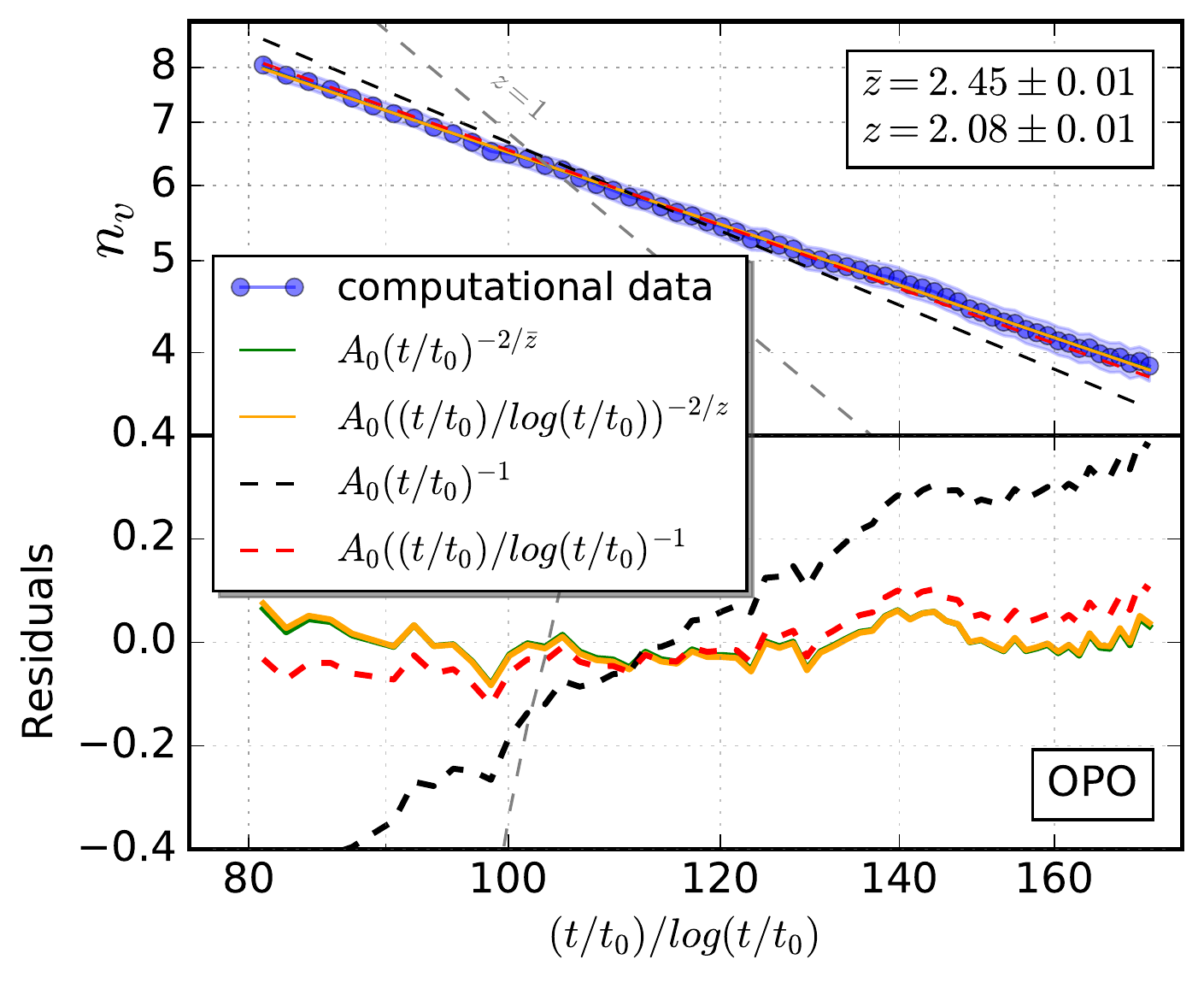}
	\includegraphics[width=\columnwidth]{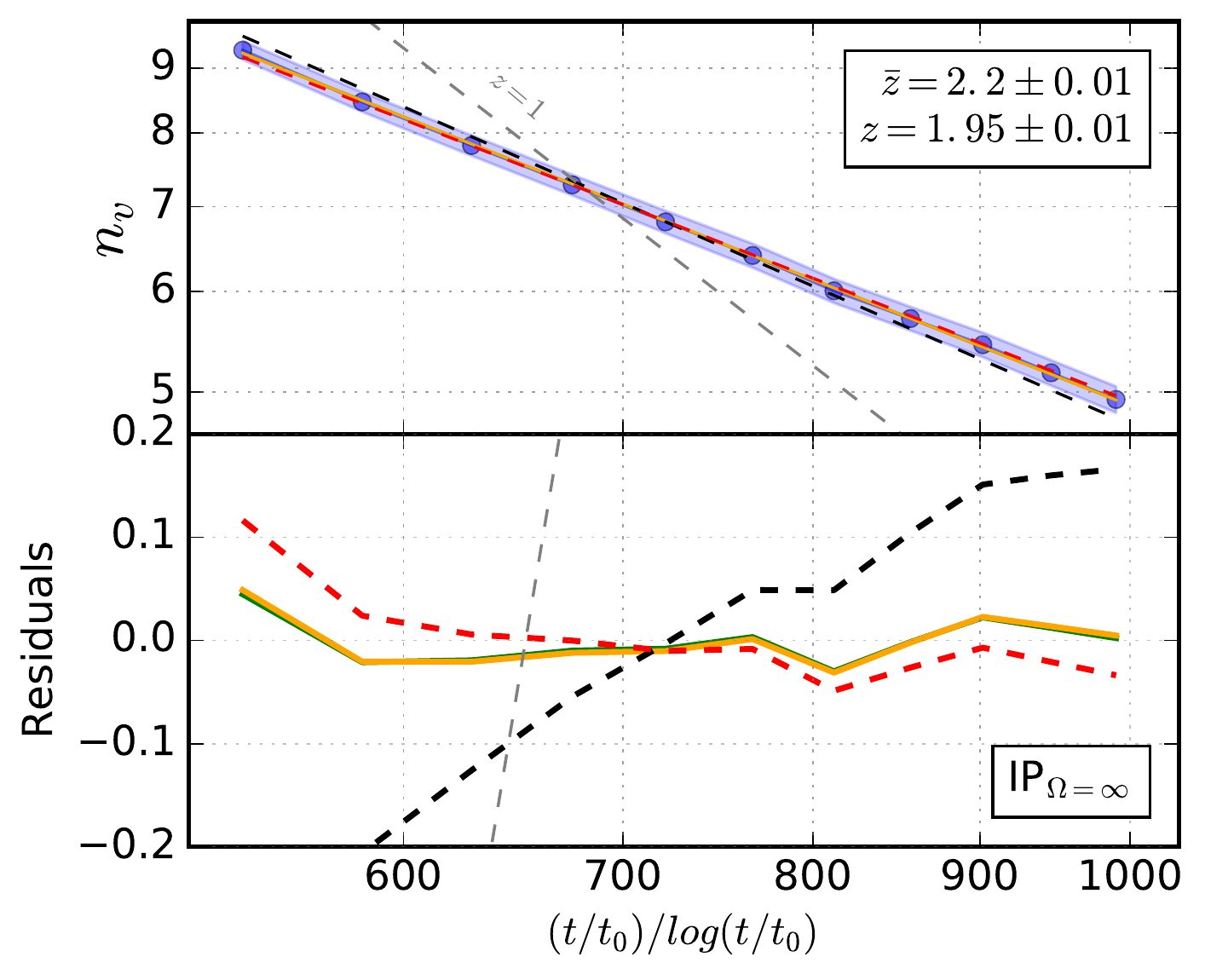}
	\includegraphics[width=\columnwidth]{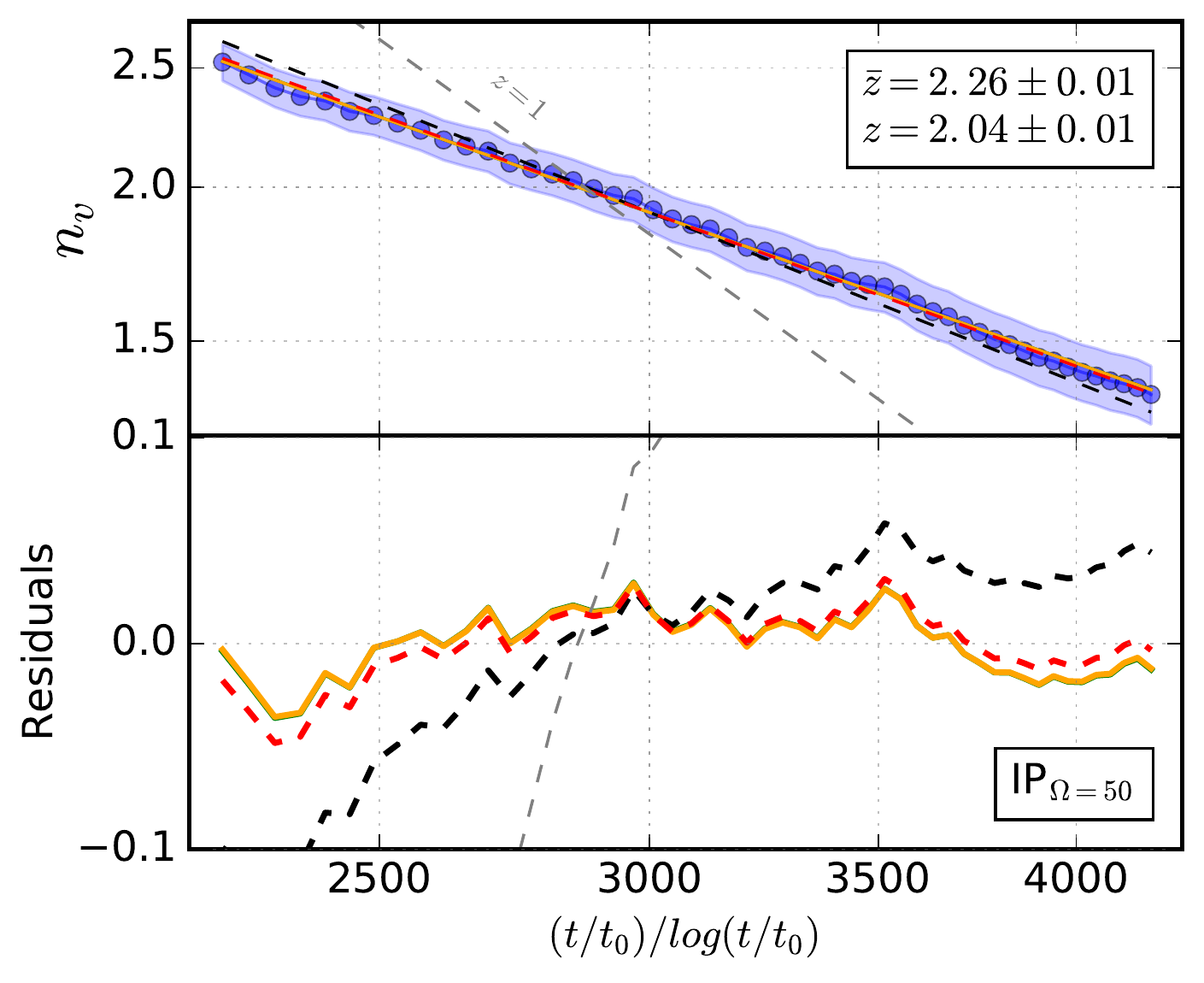}
	\caption{
		\textbf{Analysis of logarithmic corrections}. 
		Different fitting curves are used to capture the role of the logarithmic corrections appearing in the vortex number evolution at late times for the cases of OPO (top), IP$_{\Omega=\infty}$ (centre) and IP$_{\Omega=50}$ (bottom).
		Extracted exponents with ($z$) or without ($\bar{z}$) logarithmic corrections are reported in the text boxes. 
		In each case we also show the residuals of the fits.
		Parameters as in Fig.~\ref{fig:vortices_vs_time} of main paper.
	}
	\label{fig:analysys_log_correction}
\end{figure}

We now discuss the fitting criteria we adopt in the late-time stage dynamics (dark grey regions in Fig.~\ref{fig:vortices_vs_time} in main paper) of the coarsening process.
The time evolution of the average number of vortices during the annealing process is fitted with different curves;
\begin{itemize}
	\item A power law formula for a diffusive system
	\begin{equation}
	n_v(t) = A_0 \cdot (t/t_0)^{-2/\bar{z}}, \\
	\label{fitz}
	\end{equation}
	or,
	\item a function with logarithmic corrections to account for the dynamics of pairs of topological defects:
	\begin{equation}
	n_v(t) = A_0 \cdot [(t/t_0)/log(t/t_0)]^{-2/{z}},
	\label{fitz_log}
	\end{equation}
\end{itemize}
where $A_0$, $\bar{z}$ and $z$ are free parameters of the fit, and $\bar{z}$, $z$ correspond to the extracted dynamical critical exponents in the presence, or absence, of logarithmic corrections.

Constraining the dynamical exponent to the anticipated $z = 2$ value \red{\cite{Ajelic2011quench}}, we also show fits to defects dynamics with
\begin{eqnarray}
n_v(t) &=& A_0 \cdot (t/t_0)^{-1},  	\label{fitz2} \\
n_v(t) &=& A_0 \cdot [(t/t_0)/log(t/t_0)]^{-1} 	\label{fitz2_log}\;.
\end{eqnarray}
Comparison of different fitting curves are shown in Fig.~\ref{fig:analysys_log_correction} for OPO (top), IP$_{\Omega = \infty}$ (centre) and IP$_{\Omega = 50}$ (bottom) cases within our region of convergence (grey band), from which we can infer numerous conclusions. 

Firstly, while to lowest order all fits of Eqs. (\ref{fitz})-(\ref{fitz_log}) provide a reasonable description, careful consideration rules out the value $z=2$ in the absence of logarithmic corrections (black dashed lines).
Moreover, although fits of Eqs. (\ref{fitz}) and (\ref{fitz_log}) (solid green and yellow lines respectively) are practically indistinguishable (even in terms of their residuals!), we stress that the exponent ${z}$ extracted by fitting Eq.~(\ref{fitz_log}) [with logarithmic corrections] is always closer to the anticipated theoretical value for the dynamical exponent ($z=2$) than the extracted exponent $\bar{z}$ obtained from the fit based on Eq.~(\ref{fitz}).
This suggests that, to the extent that the dynamical exponent should be consistent with a value of 2, logarithmic corrections {\em have to} be present in the system, thus confirming the theoretically-anticipated description~ \red{\cite{Abray2000breakdown,Ajelic2011quench}}.
A detailed analysis on the evolution of these exponents on the chosen evolution time window is given below.

However, we should already highlight here that fitting our data with the ``conservative superfluid'' prediction of $z=1$ \red{\cite{Akulczykowski2017phase,Adamle1996phase,Awilliamson2016universal}} based on 
\begin{equation}
n_v(t) = A_0 \cdot [(t/t_0)/log(t/t_0)]^{-2},
\end{equation}
reveals strong disagreement for all three cases (as shown by dashed grey lines in Fig.~\ref{fig:analysys_log_correction}), thus ruling out such a value (in our chosen late-time windows).

To further highlight the importance of logarithmic corrections and the unequivocal observation of the correct universal dynamics at appropriately late-time evolution, we next analyse the temporal evolution of the numerically-extracted dynamical exponent $z$ obtained both from the characteristic length $L(t)$ (through analysis of the collapse in correlation function $g^{(1)}/g^{(1)}_{SS}$) and from the average number of vortices $n_v(t)$.

\begin{figure}[t]
	\includegraphics[width=1\columnwidth]{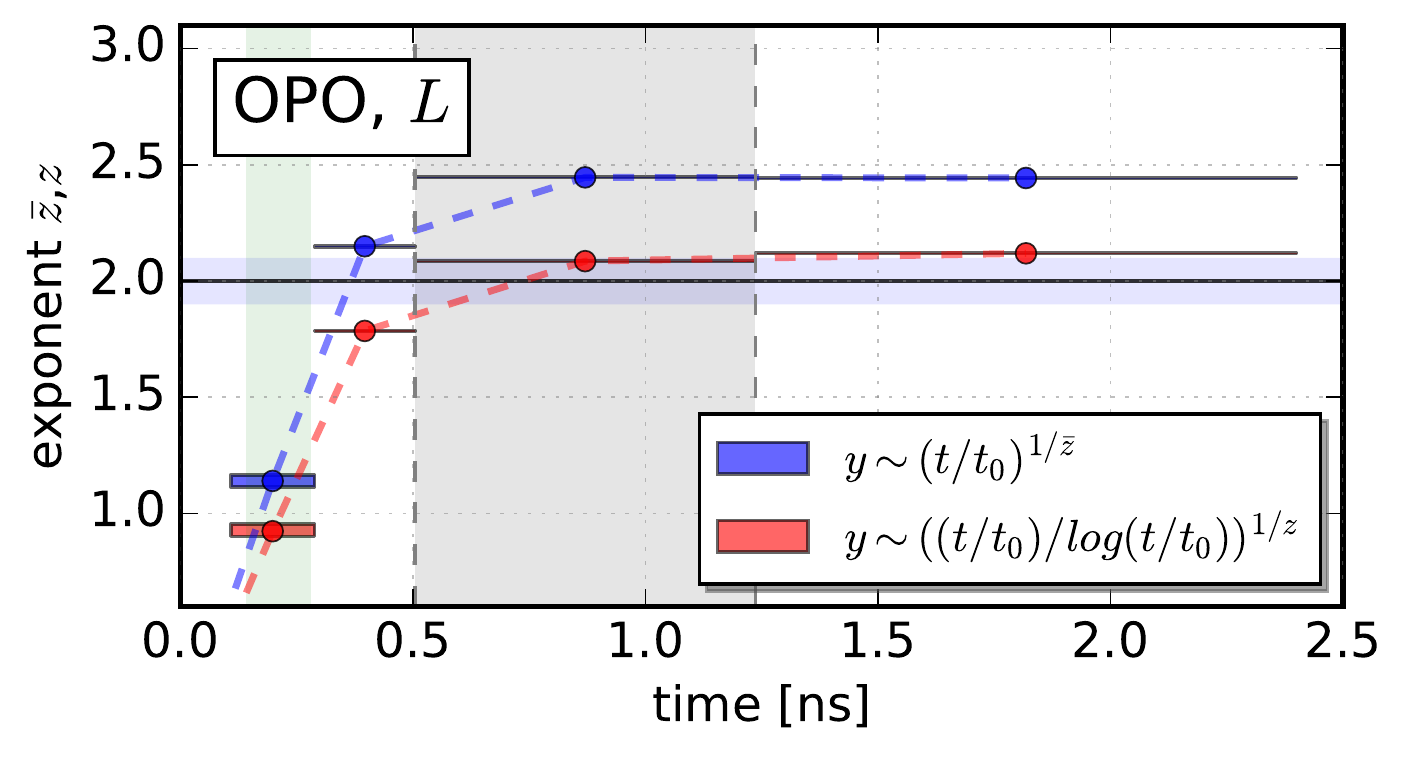}
	\includegraphics[width=1\columnwidth]{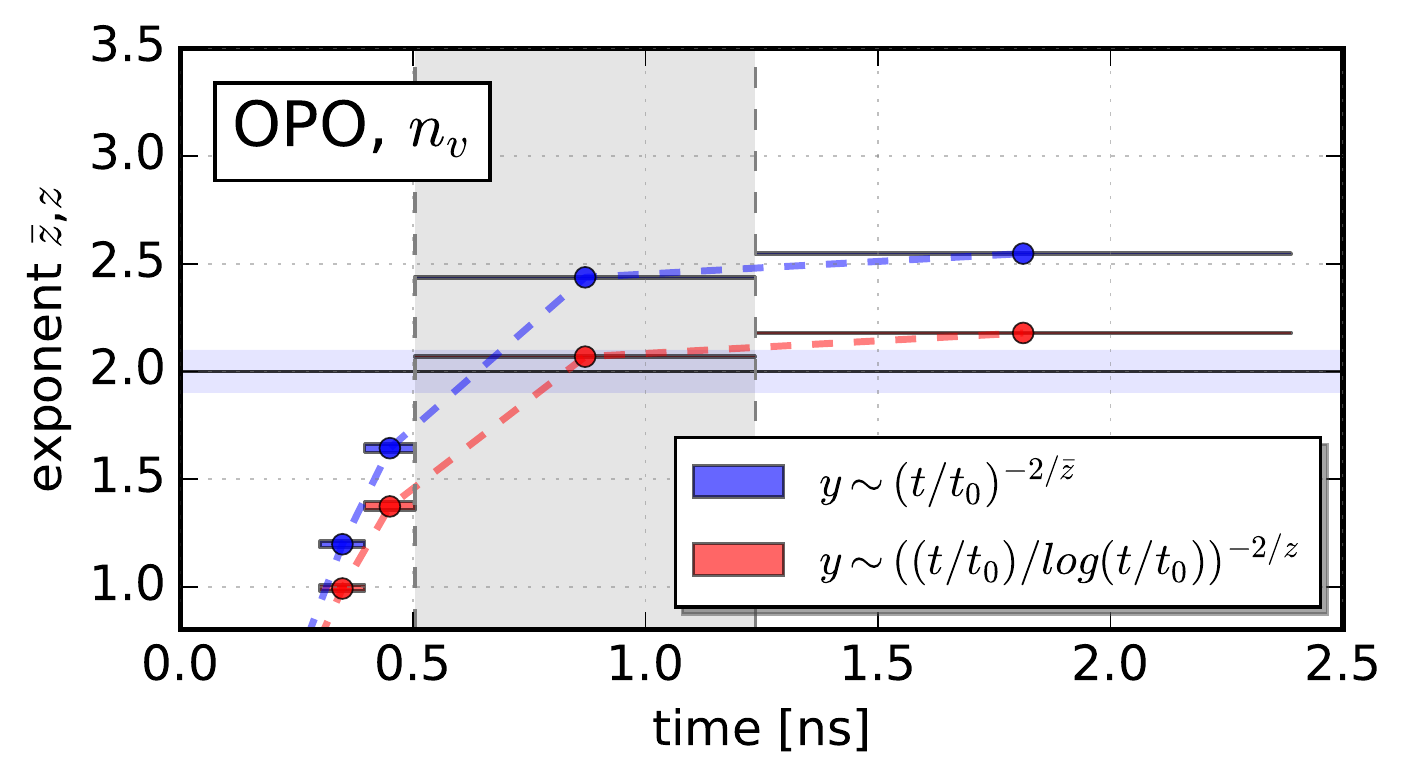}			
	\caption{
		\textbf{Time evolution of the exponent $z$ for OPO-regime.} Exponents $z$ (circles) are extracted as a function of different temporal windows, for both characteristic length (top) and number of vortices (bottom).
		Results in absence (blue) and presence (red) of logarithmic corrections are compared.
		Temporal averaging intervals and numerical-fit errors are reported for completeness by the bands around the points. 
		{The vertical green and grey bands indicate corresponding regions shown in Figs. \ref{fig:g1_collapse} and \ref{fig:vortices_vs_time} of the main paper}.
	}
	\label{fig:z_evolution_OPO}
\end{figure}
\begin{figure}
	\includegraphics[width=1\columnwidth]{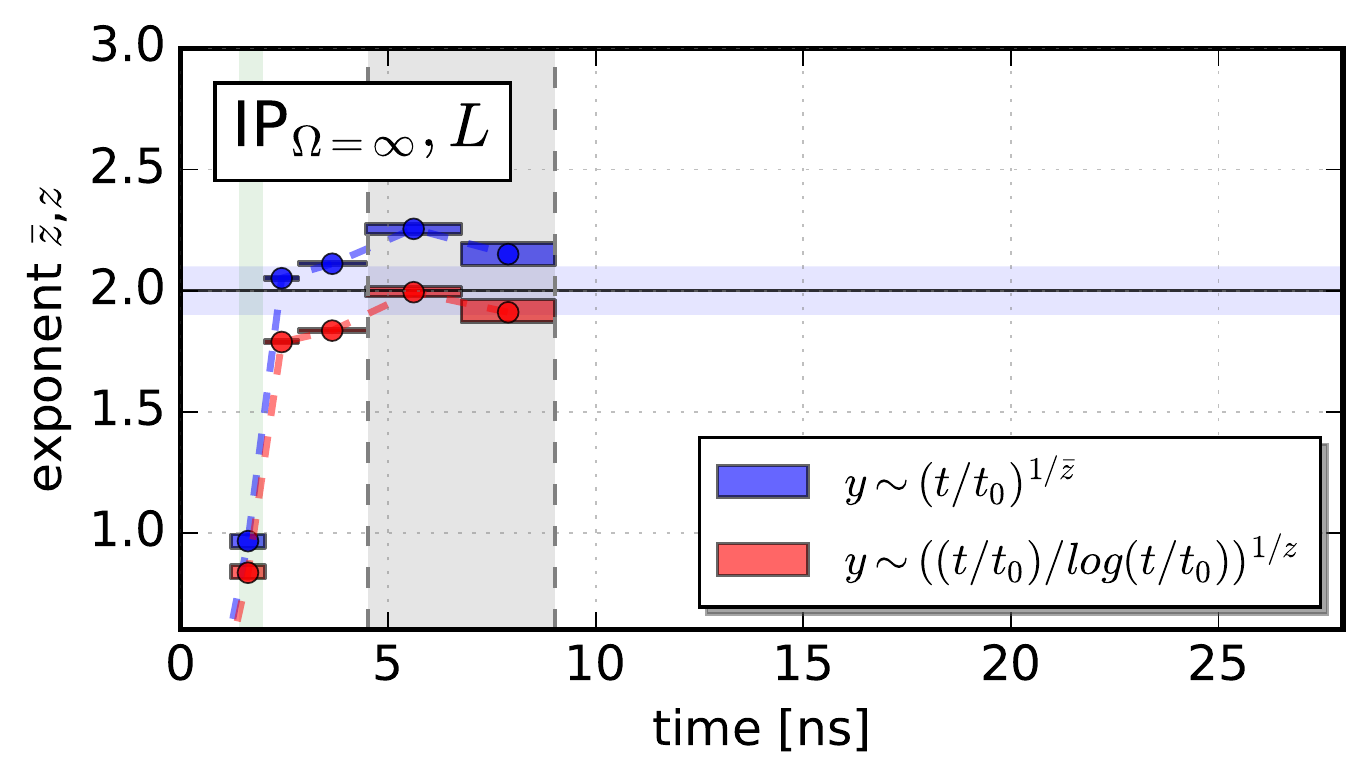}
	\includegraphics[width=1\columnwidth]{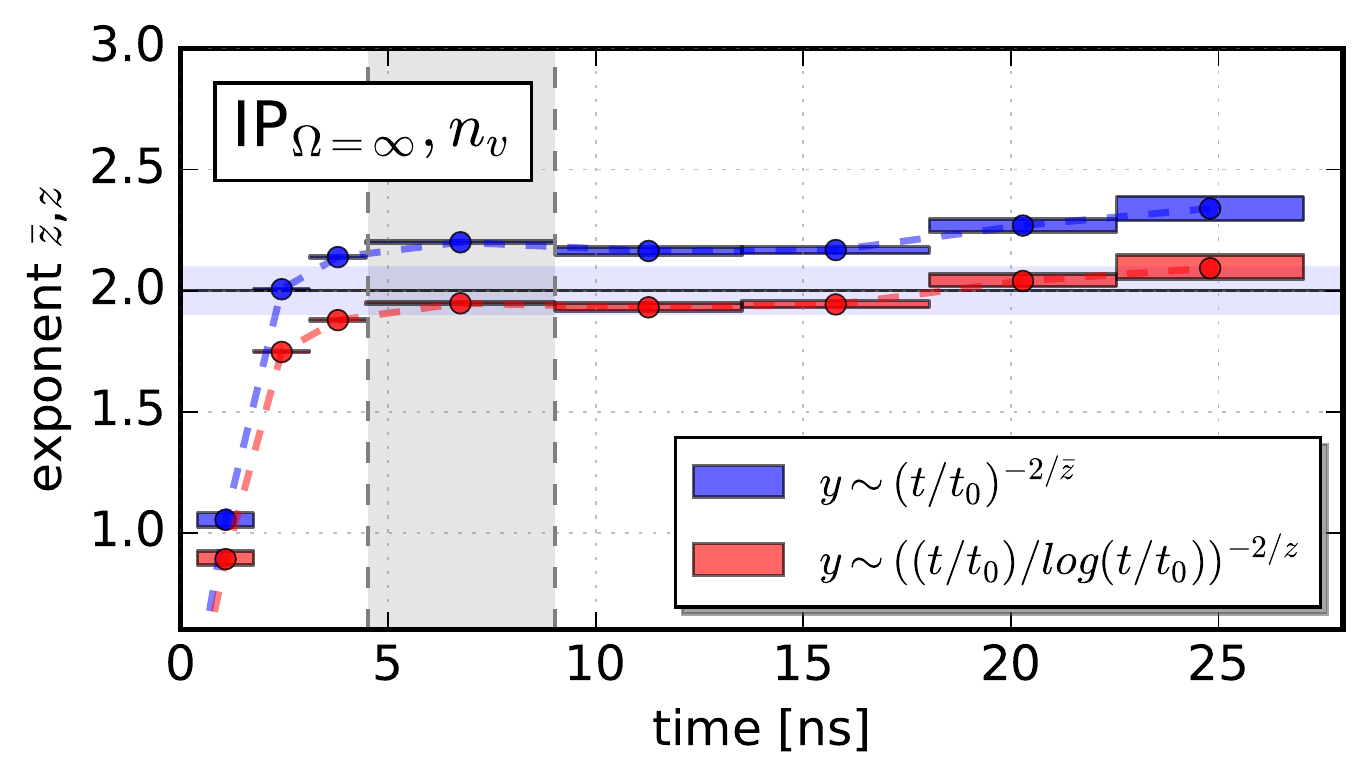}		
	\caption{
		Same as Fig.~\ref{fig:z_evolution_OPO} but for frequency-independent pumping scheme.
	}
	\label{fig:z_evolution_IPinf}
\end{figure}
\begin{figure}
	\includegraphics[width=1\columnwidth]{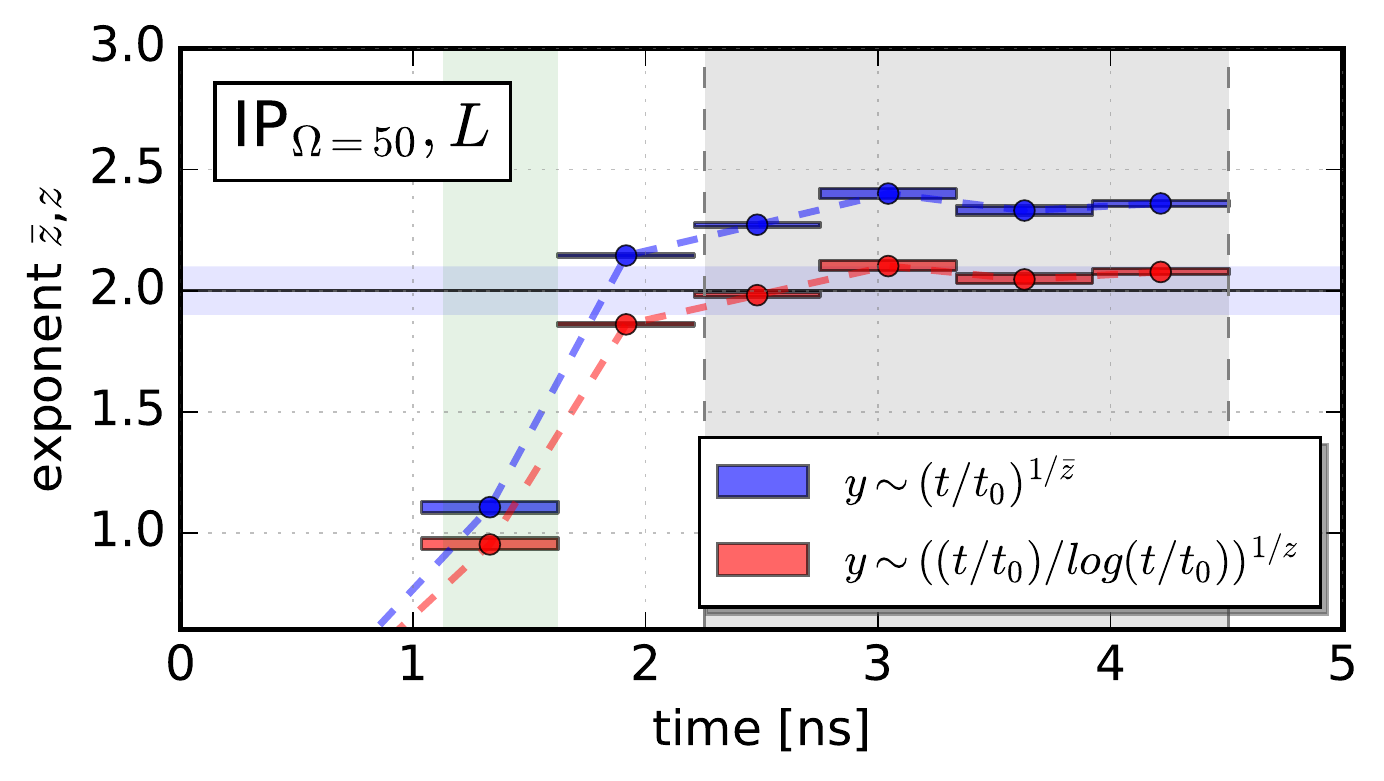}
	\includegraphics[width=1\columnwidth]{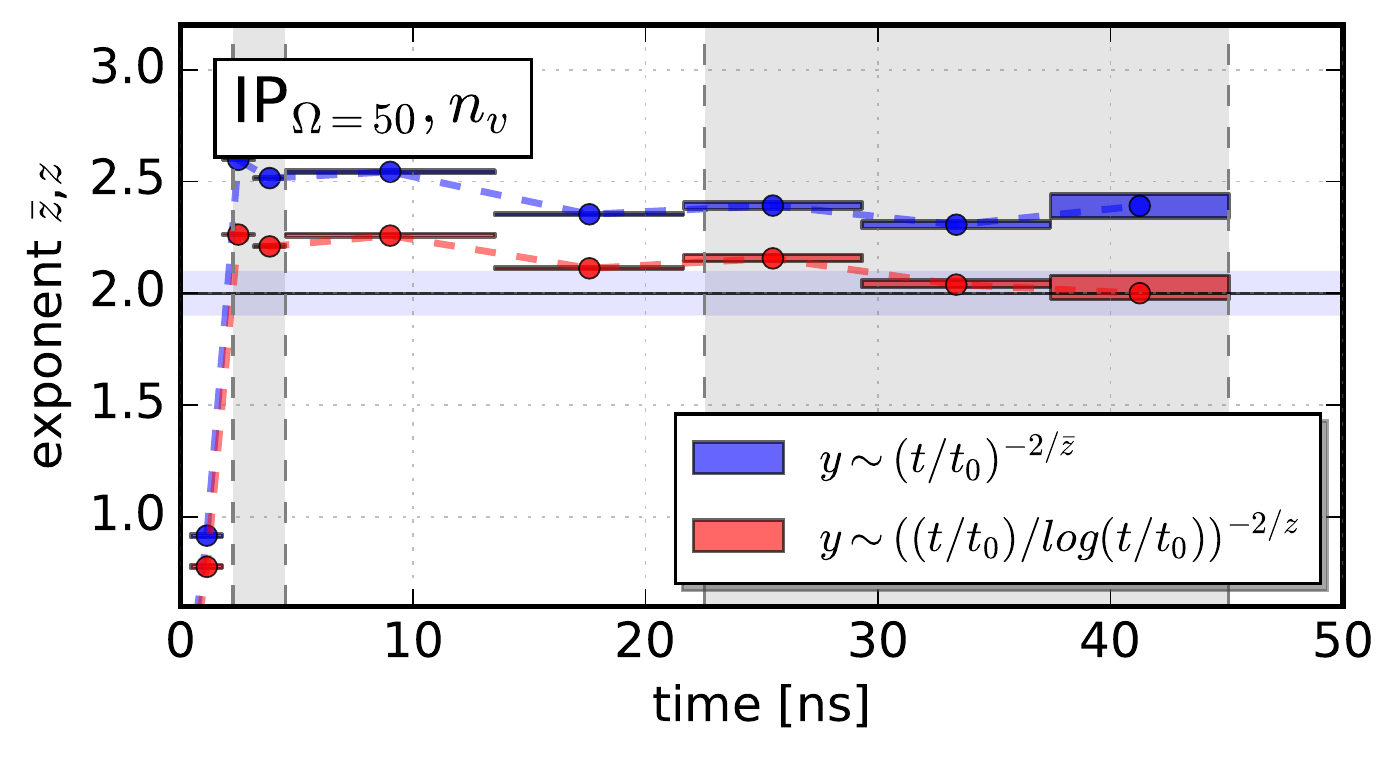}		
	\caption{
		Same as Fig.~\ref{fig:z_evolution_OPO} but for frequency-dependent pumping scheme. 
		Note the different time axis in this case associated with the disparity in time scales for collective and topological  channel equilibration.
		See Fig. 3 (bottom) in main text. 
	}
	\label{fig:z_evolution_IP}
\end{figure}

The dynamical exponents are extracted from the number of vortices by fitting our data with Eqs. (\ref{fitz}) and (\ref{fitz_log}) as above.
For the corresponding determination based on the characteristic length, $L(t)$, we first obtain $L(t)$ from our simulations by intersecting the graph of $g^{(1)}/g^{(1)}_{SS}(L(t))$ at a value of $\eta = 0.5$, ensuring we are insensitive to both short-range and finite-size effects. We then fit the extracted $L(t)$ by the corresponding relations \red{\cite{Abray2000breakdown,Ajelic2011quench}}
\begin{eqnarray}
L(t) &=& B_0 \cdot (t/t_0)^{1/\bar{z}},  	\label{fitL} \\
L(t) &=& B_0 \cdot [(t/t_0)/log(t/t_0)]^{1/{z}} 	\label{fitzL_log},
\end{eqnarray}
where $B_0$, $\bar{z}$ and $z$ are free parameters.
The time evolution of the numerically-extracted dynamical exponents are shown in Figs. \ref{fig:z_evolution_OPO}-\ref{fig:z_evolution_IP} for both cases of vortex (top) and length scale (bottom) determination, with extracted average exponents indicated by circles; the widths and heights of the ``bands'' around such points correspond respectively to the temporal averaging intervals and the error bars from the numerical fits.
The vertical green and grey bands indicate corresponding regions shown in Figs. \ref{fig:vortices_vs_time} of the main paper.

From such plots, corresponding to different parameter space and pumping schemes, we can infer the following general conclusion about the evolution of the exponent $z$:
\begin{itemize}
	\item{{\bf Early time dynamics with $z \approx 1 \ll 2$}}:
	At early times, the system is rapidly approaching (and eventually crossing) the critical region. The dynamical exponents from both characteristic length $L$ and vortex number $n_v$ agree, but scaling hypothesis does not hold in this regime {(i.e. the correlation function do not collapse onto each other)}. Within this early-stage evolution, it is possible to identify a precise time-window where $z \sim 1$ (shown by the vertical light green bands in Figs. \ref{fig:z_evolution_OPO}-\ref{fig:z_evolution_IP}).
	\item {\bf Intermediate dynamics with $z \approx 2$ but no discernible logarithmic corrections:}
	During this stage, the system has crossed the transition but not relaxed enough for logarithmic corrections to become visible. Both functions are well-fitted by Eqs. (\ref{fitz}) and (\ref{fitL}), yielding a dynamical exponent $z\sim2$ which do not include logarithmic correction. 
	\item {\bf Late-time dynamics with $z \approx 2$ with evident presence of  logarithmic corrections:}
	After sufficient evolution, the system enters the regime where the scaling hypothesis holds: in this regime the logarithmic corrections are measurable and the dynamical exponent $z\sim2$ is consistent with fitting curves with Eqs. (\ref{fitz_log}) and (\ref{fitzL_log}). 
	{At such times, the vortex decay has slowed down, with the system left with vortex-antivortex pairs only}.
\end{itemize}

We stress that in all cases, the exponents extracted at late times by fittings curves with logarithmic corrections are much closer to the expected value of $z=2$ than corresponding fits without logarithmic corrections. As a guide to the eye, Figs. \ref{fig:z_evolution_OPO}-\ref{fig:z_evolution_IP} highlight the interval $1.9<z<2.1$ by a horizontal light-blue shaded region.

Summarising, we note that our detailed analysis demonstrates that late-time dynamics of parametrically and incoherently pumped exciton-polariton systems follow a unique scaling law, consistent with that theoretically predicted for two-dimensional geometries in the context of the 2D XY model \red{\cite{Abray2000breakdown,Ajelic2011quench}}. We also stress that achieving the correct dynamical exponent $z=2$ requires very extensive numerical simulations featuring long-time evolution (such that the system fully enters the correlation-function collapse window where the scaling hypothesis holds), a very high temporal resolution during all dynamics and a large number of independent numerical realisations.

\section{Self-consistency of characteristic length-scale} 
\label{app:comparation_density_scale}

\begin{figure}
	\includegraphics[width=1\columnwidth]{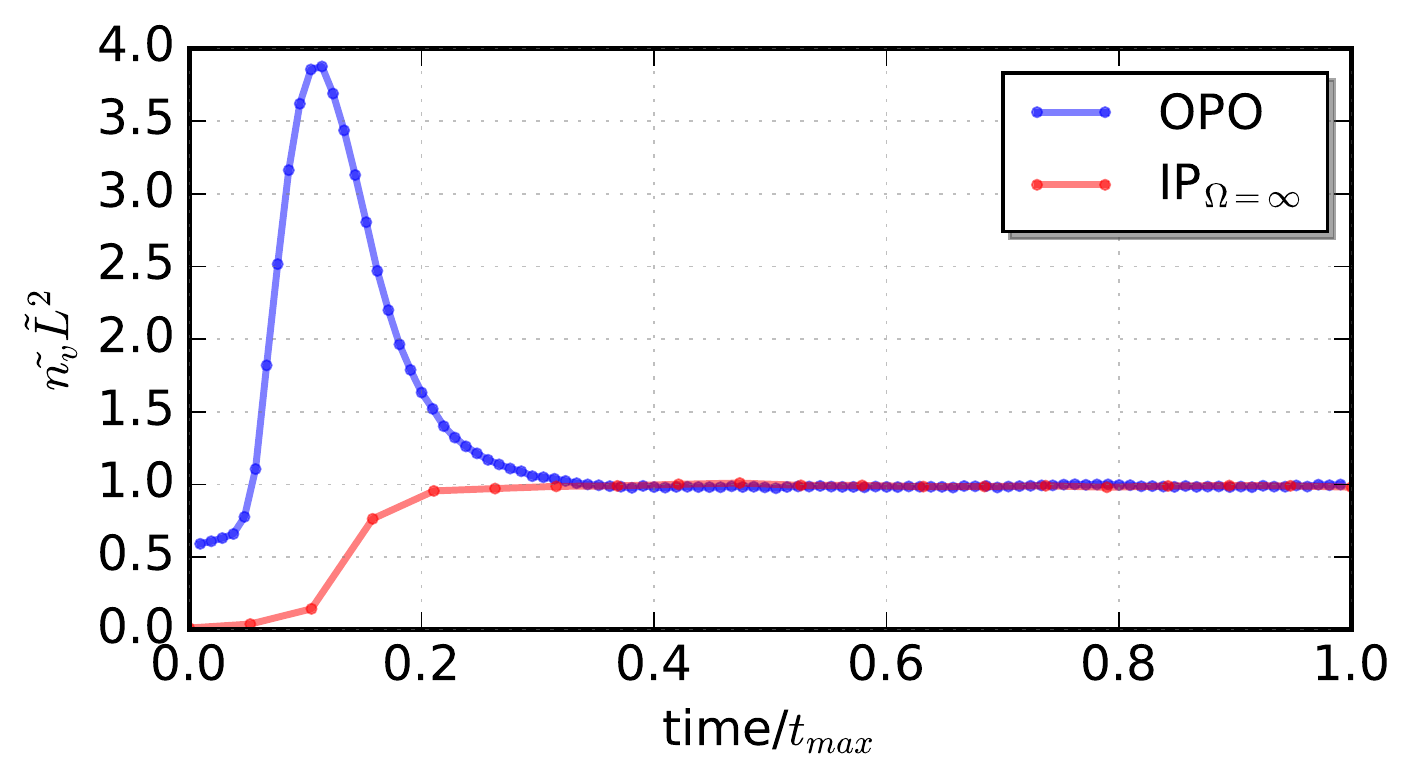}
	\caption{
		\textbf{Comparison of the number of vortices with the characteristic length scale of the system}. Scaled number of topological defects  $\tilde{n_v}=n_v(t)/n_v(t_{max})$ times the scaled characteristic length $\tilde{L}=L(t)/L(t_{max})$ squared as a function of time for the parametrically (blue) and incoherently pumped (red) system.
		We show results from the start of the quench ($t=0$) to the end of the collapse region ($t=t_{max}$).
		We observe that at late-time dynamics the system fulfils $\tilde{n_v} \tilde{L}^{2}~\sim~1$, indicating that during dynamical scaling the mean domain size $L_v \sim n_v^{-1/2}$ is proportional to the length scale $L(t)$.
	}
	\label{fig:relation_nv_L}
\end{figure}

The scaling hypothesis implies the existence of a unique length scale $L(t)$. 
This length scale extracted from the collapsing correlation functions should agree with the corresponding length scale $L_v(t)$ obtained from the vortex density through $L_v(t) \sim n_v(t)^{-1/2}$ which coincides to the mean distance between vortices.
Such a correspondence should happen at sufficiently late times in the system evolution.

Fig.~\ref{fig:relation_nv_L} demonstrates clearly that, at appropriately late times the scaled quantity $\tilde{n_v} \tilde{L}^{2}$, where $\tilde{n_v}~=~n_v(t)/n_v(t_{max})$ and  $\tilde{L}~=~L(t)/L(t_{max})$, converges to a constant value of one, confirming that the mean distance between vortices $L_v(t)$ is always proportional to the growing length scale $L(t)$ during the dynamic scaling.

\section{Dependence of $z_L$ on the intersection point  $\eta$}
\label{app:z_intersection_point}

\begin{figure}
	\includegraphics[width=1\columnwidth]{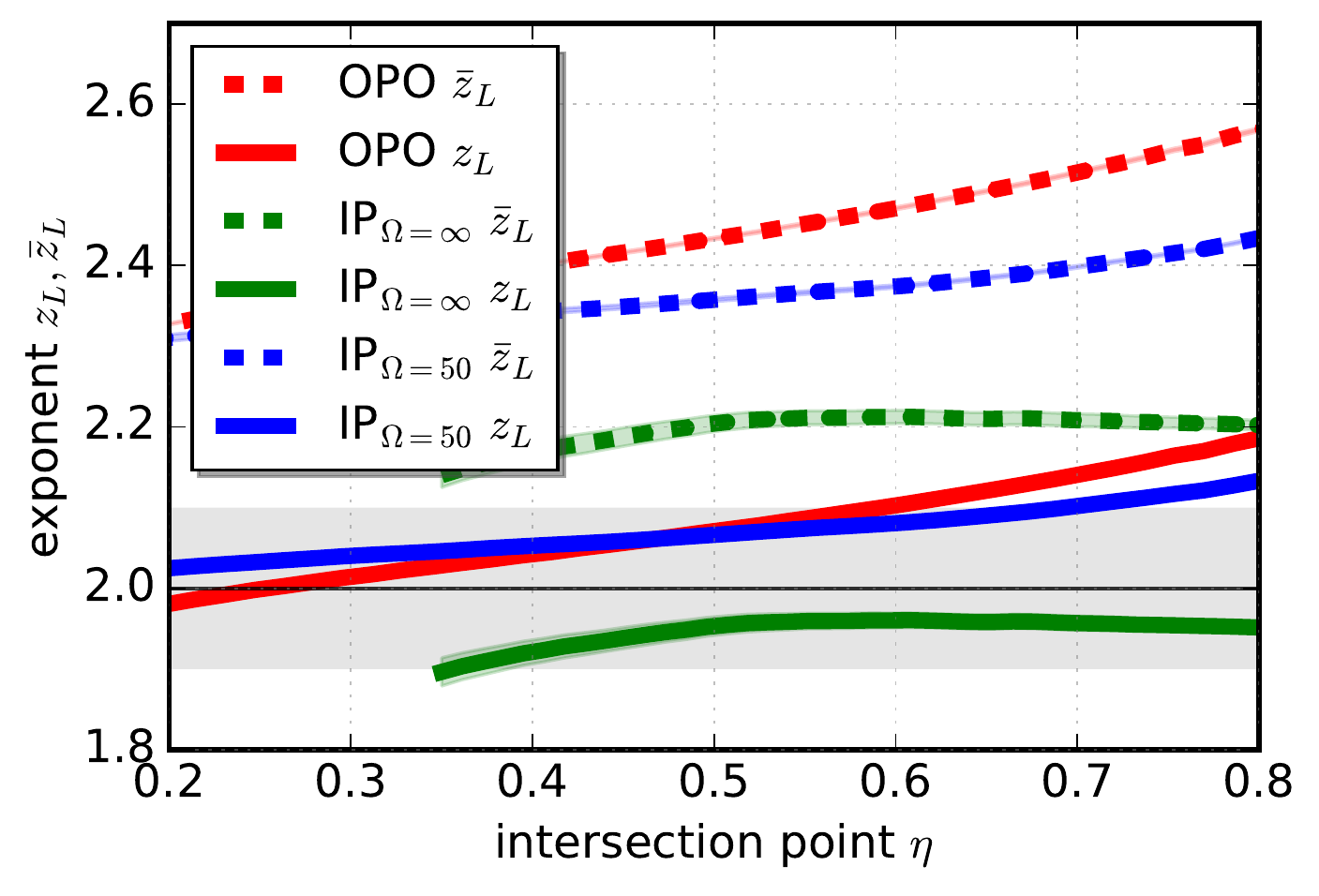}
	\caption{
		\textbf{Analysis on the dependence of $z_L$ on the intersection point  $\eta$}. We study the dependence of exponents $z$ extracted from the characteristic length scale $L(t)$ on intersection condition $(g^{(1)}/g^{(1)}_{SS})(L) = \eta$ within the collapse region for the three systems analysed: OPO (red), IP$_{\Omega = 50}$ (blue) and IP$_{\Omega = \infty}$ (green). Values of exponents $z_L$ (solid lines) are always closer to $2$ (grey band) than exponents $\bar{z}_L$ (dashed lines). 
	}
	\label{fig:z_intersection_point}
\end{figure}

In order to obtain the characteristic length-scale from the correlation function essential for demonstrating dynamical scaling, we have used the condition $ (g^{(1)}/g^{(1)}_{SS})(L) = \eta$, with the value of $\eta$ chosen as  $0.5$ (OPO, IP$_{\Omega = 50}$ and IP$_{\Omega = \infty}$).

Here we confirm the validity of our results, by a detailed consideration of the behaviour of the exponents $z_L$ and $\bar{z}_L$ extracted by fitting $L(t)$ (with curves Eqs.~(\ref{fitL}) and (\ref{fitzL_log}) respectively) over the entire time window for which the scaling hypothesis holds (corresponding to the dark grey regions of Fig.~\ref{fig:vortices_vs_time} in main paper) as a function of the intersection point $\eta$.
Such a dependence is shown in Fig.~\ref{fig:z_intersection_point}, which confirms our key point that the exponent $z_L$ obtained in the presence of logarithmic corrections is closer to the predicted value of $2$, than the numerically-extracted exponent, $\bar{z}_L$, obtained in the absence of logarithmic corrections.

Moreover, we stress that the exponent $z_L$ lies within the range $1.9 < z_L < 2.1$ for a wide range of $\eta$ values including values of $\eta$ used to extract the growing length scale $L(t)$ in previous works, i.e. $\eta = 0.25$ in Ref. \red{\cite{Akulczykowski2017phase}} and $\eta = 0.3$ in Ref.~\red{\cite{Ajelic2011quench}}.
Deviations only occur at quite extreme values of $\eta$, where finite-size and short-range effects come into play.

\section{Role of nonuniversal microscopic system time-scale $t_0$ on the dynamical exponent $z$}
\label{app:t0}

\begin{figure}
	\includegraphics[width=1\columnwidth]{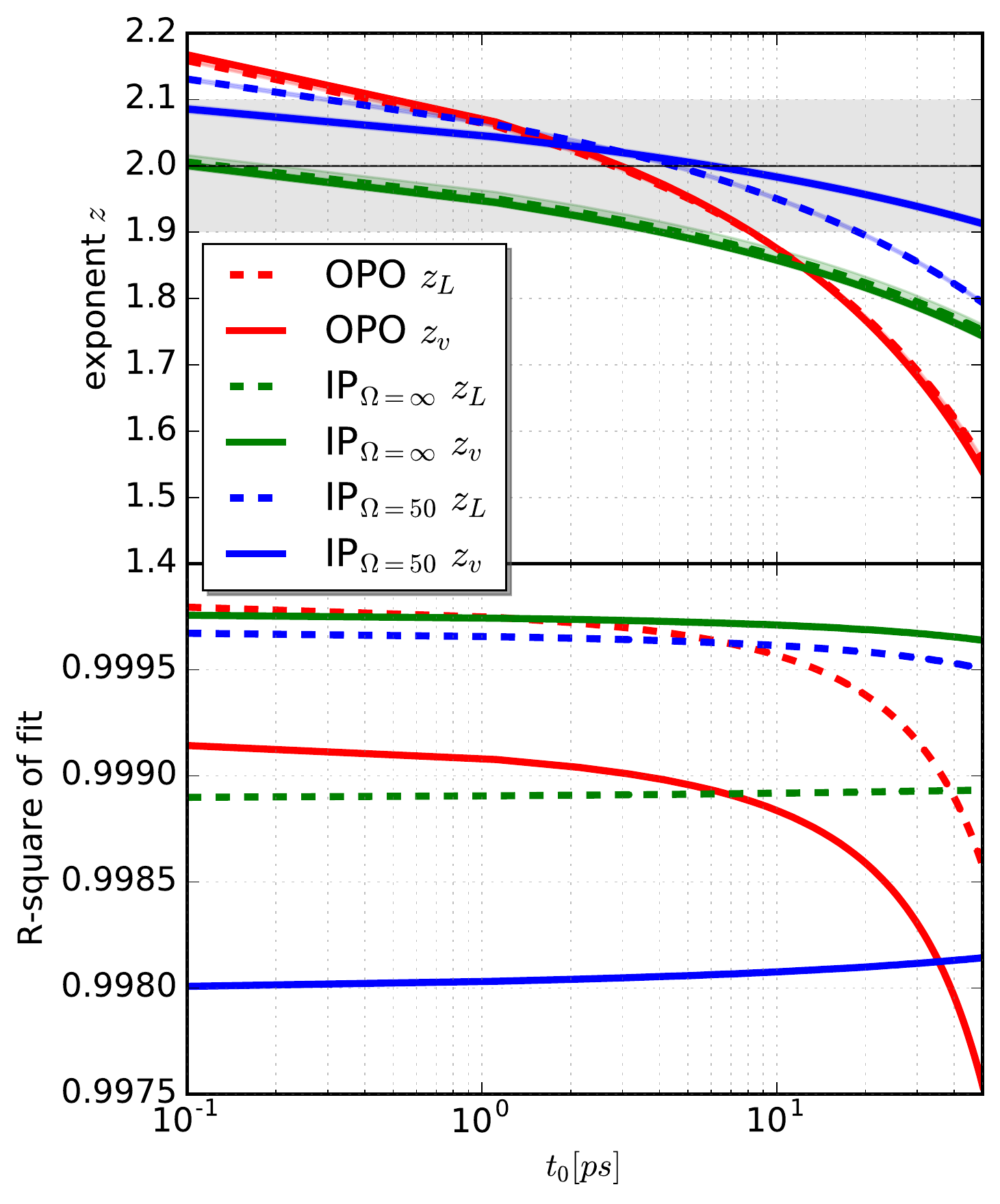}
	\caption{
		\textbf{Dependence of $z$ on the time-scale $t_0$}. 
		Dependence of exponent $z$ obtained by fitting both number of vortices and characteristic length within the collapse region for the three cases. 
	}
	\label{fig:z_time_scale}
\end{figure}

The introduction of the anticipated logarithmic corrections through the related formulas
\begin{eqnarray}
n_v(t) & = & A_0 \cdot [(t/t_0)/log(t/t_0)]^{-2/{z}} \nonumber \\
L(t) &=& B_0 \cdot [(t/t_0)/log(t/t_0)]^{1/{z}}
\end{eqnarray}
introduces a nonuniversal microscopic system timescale, $t_0$. 
{The presence of $t_0$ is typically ignored}, on the grounds that it should not affect universal features, but its detailed consideration is nonetheless relevant for the precise slope of the corresponding curves for $n_v$ and $L$.
Following common practice \cite{bray2000breakdown,jelic2011quench}, throughout this work we have effectively chosen $t_0$ such that it ``drops out'' of the corresponding equations. By considering the dependence of the universal dynamical exponent $z$ on this nonuniversal parameter, we effectively shed more light onto why such an approach is acceptable.

Since $t_0$  is a typical system timescale, it should be of broadly the same order of magnitude as all other relevant system timescales. These are listed, for both OPO and IP schemes in Table I, to demonstrate that all these timescales lie well within the range  $10^{-1}~{\rm ps}~<~\tau_{system}~<~10^1{\rm ps}$.
{This already suggests that the conventional choice of $t_0$ taking the value of 1 in the system units is acceptable, which for our current purposes would imply the already-chosen value of $t_0 = 1$ps. }
However, in order to be certain that our results do not critically depend on our choice of $t_0$, Fig.~\ref{fig:z_time_scale} displays the dependence of the dynamical exponent $z$ (in the presence of logarithmic corrections) on $t_0$ within the anticipated time window $10^{-1} {\rm ps} < t_0 < 10^1 {\rm ps}$.
From this figure it is evident that, within this range, all our numerically-extracted values for $z$ take a value $1.9 < z < 2.1$ (depicted by the horizontal light grey area in Fig.~\ref{fig:z_time_scale}).

We thus conclude that our extensive numerical analysis confirms that all three physical cases considered (OPO, IP with and without frequency-dependent pumping) are consistent with the presence of logarithmic corrections and a dynamical critical exponent of $z = 2$, as expected for the system to be in the same universality class as the 2D XY model.

\definecolor{Lightgray}{RGB}{235,235,235}

\begin{table}
	\vspace{8mm}
	\renewcommand{\arraystretch}{1.6}
	\begin{tabular}{ | c| c || c | c | c |}
		
		\toprule  
		
		\hline
		\qquad \qquad \qquad \qquad & \ \  \textbf{OPO} \ \  & \qquad\qquad\qquad & \ \  \textbf{IP$_{\Omega=50}$} \ \  & \ \  \textbf{IP$_{\Omega=\infty}$} \ \ \\ \hline \hline
		$\tau_{ph} = \tau_{ex}$ & $3.29 $ & $\tau_{LP}$ & $4.51  $ & $4.51  $ \\ 
		$\tau_{Rabi}$ & $0.15 $ & $\tau_{\gamma}$ &$4.51  $ & $4.51  $\\ 
		$\tau_{Pump}$ & $ (-)0.71 \ $ & $\tau_{P}$ &$4.78 $ & $4.96  $ \\ 
		$\tau_{g \cdot n_0}$ & $5.98  $ & $\tau_{g \cdot n_0}$  & $5.98 $ & $0.77  $ \\ 
		\hline
	\end{tabular}
	\caption{Time-scales for parametrically and incoherently pumped systems. We report photon, exciton and lower-branch polariton lifetimes ($\tau_{ph}, \tau_{ex}$ and $\tau_{LP}$ respectively), Rabi oscillation ($\tau_{Rabi}$), losses ($\tau_{\gamma}$), pumping ($\tau_{P}$) and interactions $(\tau_{g \cdot n_0})$ time-scales, in picoseconds. Parameters as in main paper).}\label{time_scales}
\end{table}


\end{document}